\documentclass[journal,twocolumn,10pt]{IEEEtran} % double column submit

\usepackage{cite}
\ifCLASSINFOpdf
  \usepackage[pdftex]{graphicx}
  \usepackage{epstopdf} % remove this
\else
\fi
\usepackage{amsmath}
\usepackage{algorithmic}
\usepackage[linesnumbered,ruled,lined]{algorithm2e}

\usepackage{amsfonts}
\usepackage{amssymb}
\usepackage{dsfont}
  \usepackage[caption=false,font=footnotesize]{subfig}
\usepackage{url}
\usepackage{array}

\usepackage{amsthm}
\usepackage{amsfonts}

\newtheoremstyle{myremark}% name of the style to be used
{}% measure of space to leave above the theorem. E.g.: 3pt
{}% measure of space to leave below the theorem. E.g.: 3pt
{}% name of font to use in the body of the theorem
{0pt}% measure of space to indent
{\bfseries}% name of head font
{.}% punctuation between head and body
{ }% space after theorem head; " " = normal interword space
{\thmname{#1}\thmnumber{ #2}: \thmnote{#3}}

\theoremstyle{theoremdd}

\theoremstyle{myremark}

\newtheoremstyle{myshortremark}% name of the style to be used
{}% measure of space to leave above the theorem. E.g.: 3pt
{}% measure of space to leave below the theorem. E.g.: 3pt
{}% name of font to use in the body of the theorem
{0pt}% measure of space to indent
{\bfseries}% name of head font
{.}% punctuation between head and body
{ }% space after theorem head; " " = normal interword space
{\thmname{#1}\thmnumber{ #2}: \thmnote{#3}}

\theoremstyle{myshortremark}
\newcounter{take}
\newtheorem{takeaway}[take]{Takeaway}

\newtheoremstyle{mydefinition}% name of the style to be used
{}% measure of space to leave above the theorem. E.g.: 3pt
{}% measure of space to leave below the theorem. E.g.: 3pt
{}% name of font to use in the body of the theorem
{0pt}% measure of space to indent
{\bfseries}% name of head font
{.}% punctuation between head and body
{ }% space after theorem head; " " = normal interword space
{\thmname{#1}\thmnumber{ #2}: \thmnote{#3}}
\theoremstyle{mydefinition}

\newcommand{\comment}[1]{{}} % used to comment stuff out easily

\usepackage{amssymb}
\usepackage{xspace}
\usepackage{bm}
\usepackage{bbm}

\def\E{\mathbb{E}}

\def\b0{{\mathbf{0}}}

\def\bI{{\mathbf{I}}}

\def\bX{{\mathbf{X}}}
\def\bY{{\mathbf{Y}}}

\renewcommand{\vec}[1]{{\mathbf{#1}}\xspace} % vectors

\newcommand{\parens}[1]{{\left(#1\right)}\xspace}
\newcommand{\brackets}[1]{{\left[#1\right]}\xspace}
\newcommand{\braces}[1]{{\left\{#1\right\}}\xspace}

\newcommand{\logten}[1]{\ensuremath{\mathrm{log}_{10}\parens{#1}}}

\DeclareMathOperator*{\argmax}{argmax}

\newcommand{\st}{\ensuremath{\mathrm{s.t.~}}\xspace}
\newcommand{\opt}{\ensuremath{^{\star}}\xspace}

\newcommand{\ind}[1]{\ensuremath{\mathbbm{1}\brackets{#1}}\xspace}

\newcommand{\todB}[1]{\ensuremath{\brackets{#1}_{\mathrm{dB}}}}

\newcommand{\msnr}{\ensuremath{\mathsf{SNR}}\xspace}

\newcommand{\msinr}{\ensuremath{\mathsf{SINR}}\xspace}

\newcommand{\minr}{\ensuremath{\mathsf{INR}}\xspace}

\newcommand{\minrth}{\ensuremath{\mathsf{INR}_\mathrm{th}}\xspace}
\newcommand{\tminrth}{\ensuremath{\overline{\minr}_\mathrm{th}}\xspace}
\newcommand{\ttminrth}{\ensuremath{\overline{\minr}^t_\mathrm{th}}\xspace}

\newcommand{\minrmax}{\ensuremath{\mathsf{INR}_\mathrm{max}}\xspace}
\newcommand{\minrthmax}{\ensuremath{\minrth^{\mathrm{max}}}\xspace}
\newcommand{\tminrtheff}{\ensuremath{\overline{\minr}_\mathrm{th,eff}}\xspace}

\newcommand{\tminrTht}{\ensuremath{\overline{\minr}_{\Th}^t}\xspace}

\newcommand{\powertx}{{P_{\mathrm{tx}}}}
\newcommand{\powernoise}{{P_{\mathrm{n}}}}

\newcommand{\Gtx}{{G_{\mathrm{tx}}}}
\newcommand{\Grx}{{G_{\mathrm{rx}}}}

\newcommand{\numbeams}{N_{\mathrm{B}}}

\newcommand{\numcells}{N_{\mathrm{C}}}

\newcommand{\setsatp}{\mathcal{P}}
\newcommand{\setsats}{\mathcal{S}}

\newcommand{\nump}{{N_{\mathrm{P}}\xspace}}
\newcommand{\nums}{{N_{\mathrm{S}}\xspace}}
\newcommand{\numc}{{N_{\mathrm{C}}\xspace}}
\newcommand{\numg}{{N_{\mathrm{G}}\xspace}}

\newcommand{\timew}{{T_{\mathrm{w}}\xspace}}
\newcommand{\timeho}{{T_{\mathrm{h}}\xspace}}
\newcommand{\Tw}{{T_{\mathrm{w}}\xspace}}
\newcommand{\Th}{{T_{\mathrm{h}}\xspace}}

\def\sfu{{\mathtt{u}}}
\def\sfv{{\mathtt{v}}}

\def\vc{{\vec{c}}}

\def\vg{{\vec{g}}}

\def\vp{{\vec{p}}}

\def\vs{{\vec{s}}}

\def\vx{{\vec{x}}}

\usepackage{xcolor}
\newcommand{\red}[1]{\textcolor{red}{#1}}

\usepackage{xspace}
\usepackage[acronym,nogroupskip,nonumberlist,nopostdot]{glossaries}
\makeglossaries

\newacronym{snr}{SNR}{signal-to-noise ratio}
\newacronym{sinr}{SINR}{signal-to-interference-plus-noise ratio}
\newacronym{sir}{SIR}{signal-to-interference ratio}
\newacronym{inr}{INR}{interference-to-noise ratio}
\newacronym{pdf}{PDF}{probability distribution function}
\newacronym{cdf}{CDF}{cumulative distribution function}
\newacronym{leo}{LEO}{low-earth orbit}
\newacronym{ngso}{NGSO}{non-geostationary orbit}
\newacronym{gso}{GSO}{geostationary orbit}
\newacronym{meo}{MEO}{medium-earth orbit}
\newacronym{su}{SU}{secondary user}
\newacronym{frf}{FRF}{frequency reuse factor}
\newacronym{los}{LOS}{line-of-sight}
\newacronym{nlos}{NLOS}{non-line-of-sight}
\newacronym{mimo}{MIMO}{multiple-input multiple-output}
\newacronym{sr}{SR}{Shadowed Rician}
\newacronym{ssr}{SSR}{Squared Shadowed Rician}
\newacronym{5g}{5G}{fifth generation}
\newacronym{eirp}{EIRP}{effective isotropic radiated power}
\newacronym{pfd}{PFD}{power-flux density}
\newacronym{pu}{PU}{primary user}
\newacronym{fcc}{FCC}{Federal Communications Commission}
\newacronym{itu}{ITU}{International Telecommunication Union}
\newacronym{nn}{NN}{neural network}
\newcommand{\sinr}{\gls{sinr}\xspace}

\newcommand{\inr}{\gls{inr}\xspace}

\newcommand{\ngso}{\gls{ngso}\xspace}
\newcommand{\gso}{\gls{gso}\xspace}
\newcommand{\meo}{\gls{meo}\xspace}
\newcommand{\eirp}{\gls{eirp}\xspace}

\newcommand{\leo}{\gls{leo}\xspace}

\newcommand{\cdf}{\gls{cdf}\xspace}

\newcommand{\gcdf}{\gls{cdf}\xspace}

\newcommand{\gsnr}{\gls{snr}\xspace}
\newcommand{\ginr}{\gls{inr}\xspace}
\newcommand{\gsinr}{\gls{sinr}\xspace}

\newcommand{\gnn}{\glspl{nn}\xspace}

\newcommand{\secref}[1]{Section~\ref{#1}}

\newcommand{\tabref}[1]{Table~\ref{#1}}
\newcommand{\figref}[1]{\figurename~\ref{#1}}

\usepackage{mathtools}
\usepackage{multirow}
\newcounter{mytempeqncnt}
\begin{document}


%\vspace{-0.2cm}
\title{Satellite Selection for In-Band \\Coexistence of Dense LEO Networks}

\author{%
    Eunsun Kim,~%
    Ian~P.~Roberts,~%
    Taekyun Lee,~%
	and Jeffrey~G.~Andrews%
  %  \thanks{E.~Kim and J.~G.~Andrews are with the 6G@UT Research Center and the Wireless Networking and Communications Group at the University of Texas at Austin. I.~P.~Roberts is with the Wireless Lab at UCLA.}
%    % \thanks{Updated: \today.}
%	% \thanks{E.~Kim and J.~G.~Andrews are with the 6G@UT Research Center and the Wireless Networking and Communications Group at the University of Texas at Austin, Austin, TX 78712 USA. I.~P.~Roberts is with the Wireless Lab at UCLA. Updated: \today.}
	%\thanks{%Manuscript received April 22, 2024; revised September 26, 2024; accepted November 26, 2024. 
		%This work was supported by Amazon Kuiper, 6G@UT, the Truchard Family Endowed Chair in Engineering, and the National Science Foundation Graduate Research Fellowship Program under Grant DGE-1610403.
		%The work of I.~P.~Roberts was supported in part by the National Science Foundation under Grant 1610403. 
		%Corresponding author: Eunsun Kim (esunkim@utexas.edu).}%
	\thanks{E.~Kim, T.~Lee, and J.~G.~Andrews are with the 6G@UT Research Center, Wireless Networking and Communications Group, University of Texas at Austin. I.~P.~Roberts is with the Wireless Lab, Department of Electrical and Computer Engineering, UCLA.}
    % , Austin, TX 78712 USA
    % , Los Angeles, CA 90095 USA
\vspace{-0.1cm}
}

\maketitle

\begin{abstract}  
We study spectrum sharing between two dense \leo satellite constellations, an incumbent \emph{primary} system and a secondary system that must respect interference protection constraints on the primary system. 
In particular, we propose a secondary satellite selection framework and algorithm that maximizes capacity while guaranteeing that the time-average interference and absolute interference inflicted upon each primary ground user never exceeds specified thresholds.
We solve this NP-hard constrained, combinatorial satellite selection problem through Lagrangian relaxation to decompose it into simpler problems which can then be solved through subgradient methods.
A high-fidelity simulation is developed based on public FCC filings and technical specifications of the Starlink and Kuiper systems.  
We use this case study to illustrate the effectiveness of our approach and that explicit protection is indeed necessary for healthy coexistence. 
%Compared to a conventional strict protection constraint, which leads to heavy under-utilization of the secondary system, our proposed interference protection constraint provides much-needed flexibility in secondary satellite selection without substantially compromising protection. 
%This results in secondary utilization on the order of 70--100\% while exhibiting interference violation rates of at worst 10\%, for instance. 
We further demonstrate that deep learning models can be used to predict the primary satellite system associations, which helps the secondary system avoid inflicting excessive interference and maximize its own capacity.

\end{abstract}

\glsresetall % resets glossary terms after abstract, comment out if desired
%\newpageth
%\tableofcontents

%\vspace{-0.15cm}
\section{Introduction}

Dense \leo satellite constellations are poised to play a vital role in delivering seamless global broadband wireless services across the globe \cite{sat_integ, sat_6g, path_6g, daesik_sat}. With Starlink, a prominent frontrunner, already delivering commercial services using more than 6,000 satellites in orbit \cite{sat_stats}, the potential of \leo networks to bridge connectivity gaps and provide universal broadband access is increasingly evident. 
In addition to Starlink, other companies like OneWeb, AST Spacemobile, Amazon, and other emerging players are actively advancing \leo satellite communication systems for a variety of use cases including direct-to-handset. %As these deployments continue, these systems face the demanding challenge of sharing limited spectrum resources, requiring careful planning, coordination and innovation to ensure efficient coexistence and maximize service delivery in an increasingly active orbital environment \cite{47cfr25_261, itur1323,fnprm}.

To facilitate fair and efficient spectrum sharing, the \gls{fcc} gives precedence (or incumbency) to LEO systems which applied for launch rights in earlier so-called processing rounds, referred to as \textit{primary} systems, than others \cite{fnprm}, termed \textit{secondary} systems. 
Consequently, it is the expectation of the \gls{fcc} that each system either {coordinates} with or {protects} systems which acquired launch rights in earlier processing rounds \cite{fnprm}.
This work examines how two dense \leo satellite systems---one the primary and the other secondary---can coexist in the same frequency band under such expectations. 
Specifically, we investigate practical mechanisms for the secondary system to determine its satellite-to-ground cell associations, choosing which satellite should serve each ground cell. 
The goal is to optimize the secondary system's performance while ensuring it does not inflict excessive interference onto ground users of the primary system. %the independently operating primary system. 

% \vspace{-0.1cm}

\subsection{Background and Related Work}

The principle of spectrum sharing is concerned with so-called {secondary} systems not inflicting significant interference onto {primary} (or incumbent) systems when attempting to operate in some portion of frequency spectrum \cite{why_cr_satcom, ss, css}. 
Extensive research has explored various cognitive radio (CR) techniques to enable such spectrum sharing, but the practical deployment of these conventional CR paradigms remains limited. 
%This has significantly impeded the real-world application of CR, ultimately resulting in the sparse adoption of CR implementations.
Spectrum sharing in \leo satellite communications introduces additional complexities due to the high mobility of satellites and substantial communication link latency \cite{3gpp38811}, leading to outdated spectrum perception and diminished effectiveness of traditional spectrum access strategies \cite{istn}. 
Despite the shortcomings of CR techniques, LEO satellite communications offer a unique opportunity for revitalizing spectrum sharing strategies. 
Several key factors contribute to this potential: 
(i)~highly directional communication links, which result in some natural orthogonality, 
(ii)~known satellite locations in space, which usually have predictable line-of-sight propagation to Earth, 
(iii)~advances in machine learning (ML) for spectrum prediction and resource optimization, and 
(iv) the absence of alternative spectrum-sharing solutions suitable for the LEO environment. 
%These factors collectively establish a foundation for developing innovative approaches, particularly those leveraging AI-driven spectrum inference, that can address the practical limitations that have constrained terrestrial CR systems.

ML-driven spectrum inference utilizes historical and real-time data to predict spectrum occupancy and environmental dynamics, enabling a transition from reactive sensing to proactive prediction and enhancing spectrum management and decision-making \cite{ss_inference}. 
ML-powered CR systems integrate two key capabilities: perception, which predicts spectrum occupancy and primary system behavior, and reconfigurability, which optimizes spectrum access and resource allocation \cite{intel_cr}. 
Most environment perception schemes in CR systems, including satellite communication systems, primarily focus on predicting spectrum occupancy---that is, determining whether the primary system is actively using the spectrum---using various neural network architectures \cite{dl_css, cm_cnn, hc_lstm, cnn_rnn}.
Understanding spectrum usage of the primary system is vital, but in satellite systems, it is especially critical to identify directions in which primary satellites and ground users steer their beams. 
This directional information and separation allows secondary systems to share frequency resources even when the primary system is active \cite{tonkin}. 
Existing studies emphasize that avoiding interference in specific angular directions---referred to as ``avoidance angles" can effectively safeguard geostationary satellite systems from harmful interference %caused by
from \ngso systems \cite{itur1526}. Additionally, this approach has demonstrated potential in mitigating interference between \ngso satellites, including those in \leo and medium-earth orbit \cite{he, tonkin, dense_leo, feasibility_journal}.

Building on spectrum prediction, reinforcement learning is often proposed for dynamic spectrum access decisions \cite{rl_qlearning, dyna_im} in relatively small and simple satellite systems. 
However, the application of reinforcement learning to larger and more complex systems is extremely difficult due to the increased state and action spaces, resulting in slower convergence and higher computational overhead \cite{drl_5g, rl_wireless}.
In dense \leo networks, the challenge extends beyond simple spectrum access decisions.
Rather, it involves determining the optimal satellite-to-ground cell associations across an entire constellation of multi-beam satellites to maximize the secondary system's spectrum utilization while guaranteeing sufficient protection of primary users---which is the focus of this work.

% ly large avoidance angles. 

%\vspace{-0.24cm}
\subsection{Contributions}

Our recent work \cite{feasibility_journal} demonstrated that in-band coexistence of two dense \leo satellite systems is in principle feasible through strategic satellite selection, meaning that the secondary system would have at least one satellite that could serve a given secondary ground user in light of the primary satellite-to-ground user association.  In other words, no matter what the primary system is doing as far as serving its ground users, the secondary system can in principle work around it, since it has at least one satellite that can avoid interfering with any hypothetical primary ground user.   While this is an encouraging result,  how to actually execute such a selection reliably and on global scale was left as an open problem.   We are unaware of any other work that solves such a problem. 

The main contribution of the present paper, therefore, is to propose a framework and practical algorithm for optimal satellite selection across a secondary constellation.  Optimal in this context being defined as maximizing the capacity of the secondary system while limiting the interference it inflicts onto primary users, meanwhile accounting for any other operational constraints. 
The interference protection constraint has often been formulated in quite a strict manner, limiting the increase in the effective temperature of a primary user's receiver to be no more than 6\%~\cite{ntia_ipc, itur1323} due to interference from the secondary system, which translates to an \ginr of at most $-12.2$~dB.

\textbf{Novel framework and formulation for LEO spectrum sharing.} %---pairing the selected satellite to a specific cluster on the ground (?).% \cite{lmckp_david, integer_prog}. \ipr{Maybe don't cite here, cite in the text?}
Our work formulates a novel interference protection constraint and shows that it offers more flexibility to the secondary system in its satellite selection, while still offering comparable levels of protection as the aforementioned strict $-12.2$~dB constraint.
We then employ this proposed constraint to formulate a satellite selection problem which aims to maximize the capacity of the secondary system while limiting the interference inflicted upon each primary ground user---both in an absolute sense and time-averaged sense.
This problem formulation and our approach to solving it are centered on grouping ground cells into clusters, each of which is served by a single satellite.  This not only offers dimensionality reduction and mathematical tractability but also aligns well with the operation of practical deployments like Starlink and Kuiper. 

% aiming to assign a serving secondary satellite to each cluster of cells to maximize total capacity while ensuring that interference to primary users remains within a predefined threshold.

% This is accomplished through
% In this work, we begin by formulating a  more flexible yet comparably protect

% The interference constraint for spectrum sharing is often quite stringent, typically limiting the increase in the effective temperature of a primary user's receiver to be no more than 6\%~\cite{ntia_ipc, itur1323} due to interference from the secondary system, which translates to an \ginr of at most $-12.2$ dB. % inflicted by secondary system onto a primary user limited to
%Our selection process considers the primary system's dynamically changing satellite-to-ground associations and ensures that the secondary system operates within the defined interference protection constraints. 

\textbf{Practical algorithm to achieve optimal satellite associations.}
We develop a practical algorithm by transforming the challenging combinatorial nature of the formulated NP-hard problem into a sequence of simpler problems via Langrangian relaxation, each of which can be solved through subgradient iterations.
In turn, our satellite selection mechanism offers computational efficiency yet remains capable of reliably protecting primary ground users from excessive interference.  %while simultaneously maximizing the capacity of the secondary system.
% reducing the interference incurred by secondary ground users. 
Conceptually, this is accomplished through strategically associating secondary cells with satellites which are spatially separated from active primary satellites serving nearby ground cells---offering protection to primary ground users while also reducing the interference received by secondary ground users from primary satellites.

\textbf{High-fidelity simulation of Starlink and Kuiper inter-operation.}  We develop a high-fidelity simulation of two prominent \leo satellite systems---Starlink as the primary system and Kuiper as the secondary system---each consisting of thousands of satellites, using publicly available data from their FCC filings \cite{spaceX_grt, kuiper}. 
To ensure regulatory compliance, we implement a power control mechanism that adjusts transmission power based on the satellite’s path distance while adhering to maximum \eirp and power flux density (PFD) limits \cite{cfr25}. 
The simulation models each system's satellite-to-cluster associations as satellites traverse their orbits, incorporating satellite handover dynamics across multiple ground cells and association policies. 
Ground cells are modeled using Earth-centered Earth-fixed (ECEF) coordinates \cite{3gpp38811}, and each satellite is equipped with multiple spot beams \cite{kuiper,leo_comparison,num_spotbeams}, following a predefined transmit beam gain pattern within the beam mask \cite{itur1528}. Each beam is allocated to a single ground cell at any given time \cite{kuiper}, utilizing a specific time and frequency resource.
%We simulate these satellite systems over a broad geographical area, spanning major cities across Texas, USA. 

{While there is an inherent trade-off between the secondary system's capacity and the level of protection afforded to the primary system, results indicate that the secondary system can coexist with the primary system using the proposed secondary satellite selection algorithm. 
Our proposed interference protection constraint introduces flexibility in secondary satellite selection while maintaining a well-controlled compromise in protection. % requirements. 
Furthermore, we demonstrate that secondary satellite selection can be effectively performed by leveraging deep learning (DL) models that learn an undisclosed underlying primary satellite association policy. These learned models can then be employed to optimize the proposed secondary satellite selection process.}

% \begin{takeaway}[Optimizing coexistence entails careful selection of a number of parameters]
% Given the complexity of this interference scenario, it is difficult to state a one-size-fits-all solution.
% With our proposed framework, however, multiple parameters can be tuned to reliably arrive at a satellite selection mechanism which meets an acceptable violation rate while maximizing utilization.
% Regulators will play a pivotal role in defining violation rates and worst-case \ginr that are acceptable in this context.
% Beyond this, the rate at which the secondary system can perform handovers will also have decisive impact on its ability to coexist with the primary system.
% \end{takeaway}

\comment{
\textbf{We formulate this secondary satellite selection problem as a mixed-integer matching problem \cite{lmckp_david, integer_prog}, aiming to assign a serving secondary satellite to each cluster of cells to maximize total capacity while ensuring that interference to primary users remains within a predefined threshold.}
%Additionally, each cluster must be associated with up to one satellite, as detailed in \secref{sec:sat_selection}.
Given the vast number of ground cells on the surface of the Earth required for global connectivity, a single satellite is assumed to be responsible for serving multiple ground cells, collectively defined as a cluster. We therefore aim to determine the optimal serving satellite for each cluster. 
To formulate the problem, we assume that the secondary system has complete knowledge of the primary system, particularly the satellite-to-ground association. 
This knowledge enables the secondary system to select a satellite that serves a given cluster while maintaining a sufficient avoidance angle from active primary satellite-to-ground links, thereby directly impacting the \gsinr of secondary users and \ginr of primary users. 
Such information is essential for ensuring effective spectrum sharing and mitigating inter-system interference.
Although this problem is a well-known NP-hard combinatorial optimization problem \cite{np_hard}, it can be effectively addressed using Lagrangian relaxation \cite{lagrangian_relax}. 
This approach involves relaxing the constraints, decomposing the problem into subproblems, and iteratively solving them using subgradient methods.

\textbf{In \secref{sec:numerical_results}, the performance of this approach is then validated through simulations, demonstrating its effectiveness under the assumed conditions.} 
To achieve this, we develop a high-fidelity simulation of two prominent \leo satellite systems---Starlink as the primary system and Kuiper as the secondary system---each consisting of thousands of satellites, using publicly available data from their FCC filings \cite{spaceX_grt, kuiper}. 
To ensure regulatory compliance, we implement a power control mechanism that adjusts transmission power based on the satellite’s path distance while adhering to maximum \eirp and power flux density (PFD) limits \cite{cfr25}. 
The simulation models each system's satellite-to-cluster associations as satellites traverse their orbits, incorporating satellite handover dynamics across multiple ground cells and association policies. 
Ground cells are modeled using Earth-centered Earth-fixed (ECEF) coordinates \cite{3gpp38811}, and each satellite is equipped with multiple spot beams \cite{kuiper,leo_comparison,num_spotbeams}, following a predefined transmit beam gain pattern within the beam mask \cite{itur1528}. Each beam is allocated to a single ground cell at any given time \cite{kuiper}, utilizing a specific time and frequency resource.
We simulate these satellite systems over a broad geographical area, spanning major cities across Texas, USA.

\textbf{We then extend our analysis to scenarios where the secondary system lacks complete knowledge of the primary system, particularly the primary satellite-to-cluster association information---essential information in the secondary satellite selection process.} 
Due to the highly dynamic nature of \leo satellite systems, acquiring such data in real-world deployments is challenging. 
To address this, we employ AI techniques to infer the primary system's satellite-to-cluster association using \gnn. 
This inferred information is then utilized to determine the optimal secondary satellite-to-cluster association and assess the coexistence of both systems.
By enabling dynamic and intelligent secondary satellite selection, this approach facilitates more practical spectrum sharing while ensuring adequate interference protection for the primary system in real-world satellite communication environments.
}

%\vspace{-0.24cm}
\comment{
\subsection{Organization}
We present the system model and key performance metrics in \secref{sec:system_model}. %, followed by the methodology for analyzing the coexistence of dense \leo satellite systems in \secref{sec:methodology}. 
In \secref{sec:sat_selection}, we propose a satellite selection mechanism for the secondary system that maximizes capacity while adhering to interference constraints, assuming complete knowledge of the primary system’s association. A comprehensive performance evaluation of the proposed mechanism through simulations is provided in \secref{sec:numerical_results}. In \secref{sec:nn}, we extend the evaluation using inferred primary system association information obtained through training. Finally, we conclude the paper in \secref{sec:conclusion}.
}

\comment{

In this context, intelligent wireless communications have emerged as a transformative approach. The aim is to equip systems with the ability to perceive and assess available resources, autonomously learn to adapt to the dynamic wireless environment, and reconfigure their operating modes to maximize the utility of available spectrum. These capabilities—perception and reconfigurability—are central to cognitive radio systems, which are designed to intelligently adapt to changing conditions. Modern machine learning techniques further enhance this adaptability, offering significant promise in optimizing system performance through advanced decision-making and prediction capabilities. However, challenges remain, including accurately modeling primary system behavior, addressing computational complexity, and ensuring robustness in highly dynamic and diverse spectrum environments.

Cognitive radio {in the context of satellite systems} has been proposed to manage and mitigate this interference via mechanisms including spectrum sensing, underlay, overlay, and database methods \cite{satcom_cr1,  database}.
The basic idea of \textit{underlay} methods is that, when primary systems are {deemed} idle, a secondary system can use the free spectrum opportunistically \cite{satcom_cr1}. %\cite{css, ss}% based on energy detection of primary signals
On the other hand, \textit{overlay} techniques allow a secondary system to use the spectrum concurrently with primary systems, assuming the secondary system does not substantially impact normal operation of the primary systems---which naturally leads to discussions and debate on {what comprises} ``acceptable" levels of interference. 
Various mechanisms along multiple dimensions have been studied to enable such overlay coexistence, with power allocation \cite{ka_coex, pc_1} and spatial domain beamforming \cite{beam1,  beam3} being two prominent proposed routes to protect primary systems. 

Regulatory bodies play a key role in establishing clear and comprehensive rules for spectrum sharing to realize coexistence between \leo satellite systems.
The \gls{fcc}, for instance, has employed an overlay coexistence paradigm, allowing secondary systems to inflict marginal interference onto primary systems, since this can facilitate more efficient and more widespread use of spectral resources \cite{ css_mb}.
Defining the level at which interference becomes prohibitive to a primary system has proven to be a complicated task involving a variety of priorities from multiple perspectives, and as a result, it has been difficult to formulate and regulate a so-called \textit{protection constraint} which the secondary system must oblige \cite{ntia_ipc}.

This protection constraint in satellite systems has been often formulated as a threshold on the interference power relative to the noise floor of the primary system (i.e., the \gls{inr}).
For instance, it has been proposed that the resulting \ginr not be more than a threshold ranging from $-6$~dB to an even stricter $-12.2$~dB, perhaps for some specified fraction of time \cite{ntia_ipc, itur1155}. %, itur1432,itur1323}.
\comment{Beyond looking purely at interference, a throughput degradation constraint has also been considered to more meaningfully quantify the impact of interference on the primary system \cite{fnprm}.
	This is naturally more difficult to quantify from the perspective of the secondary system, since it depends on the signal quality of the primary system, and thus introduces open questions on how to satisfy and evaluate such a protection constraint.}
In this work, we consider a {strict constraint which requires that the interference inflicted onto primary ground users by the secondary system not exceed a specified \ginr threshold.}

\subsubsection{Satellite Spectrum Sharing:}
	The principle of spectrum sharing is concerned with so-called \textit{secondary} systems not inflicting significant interference onto \textit{primary} (or incumbent) systems when attempting to access some portion of frequency spectrum \cite{why_cr_satcom, ss, css}.
%	\red{The secondary system is allowed to use the spectrum concurrently with primary systems, assuming the secondary system does not substantially impact normal operation of the primary systems---which naturally leads to discussions and debate on what defines an acceptable level of interference \cite{why_cr_satcom, ss, css}.}
%	{Cognitive radio in the context of satellite systems has been proposed to manage and mitigate this interference via mechanisms including spectrum sensing, underlay, overlay, and database methods \cite{satcom_cr_overlay, cr_sat1, satcom_cr1,  database}.
%	The basic idea of \textit{underlay} methods is that, when primary systems are deemed idle, a secondary system can use the free spectrum opportunistically \cite{why_cr_satcom, ss, css}. }
%	On the other hand, \textit{overlay} techniques allow a secondary system to use the spectrum concurrently with primary systems, assuming the secondary system does not substantially impact normal operation of the primary systems---which naturally leads to discussions and debate on {what defines an acceptable level of interference}. 
%	\red{Regulatory bodies play a key role in establishing clear and comprehensive rules for spectrum sharing to realize coexistence between satellite systems.
%	The \gls{fcc}, for instance, has employed an overlay coexistence paradigm, allowing secondary systems to inflict marginal interference onto primary systems, since this can facilitate more efficient and more widespread use of spectral resources \cite{why_cr_satcom, css_mb}.}	
\red{Various mechanisms have been studied to mitigate interference in spectrum sharing \cite{sharma_pc, tonkin,  beam_pointing}. %, with power allocation \cite{sharma_pc} and spatial domain beamforming \cite{tonkin,  beam_pointing} being two prominent proposed routes to protect primary systems. 
	In particular, spatial domain beamforming employing an \textit{avoidance angle} or \textit{look-aside} technique imposes a minimum separation angle between a primary satellite-user link and interfering links from secondary systems \cite{itur1526}. % such as \leo 
	These techniques were initially proposed to protect \gso systems from harmful interference caused by \ngso systems but also have promise in mitigating interference between \ngso satellites \cite{he, tonkin, dense_leo}, including \leo and \meo satellites.
	The work \cite{interference} concludes that downlink interference of early-stage \ngso satellite communication systems of the 1990s is effectively managed by high beam gain antennas.
	A notable survey of existing work on interference studies of modern satellite systems \cite{he} draws a similar conclusion that interference between satellite systems is well controlled by a high beam gain mask and an avoidance angle of at least $5^\circ$,}  \red{but this conclusion should be re-evaluated in the context of emerging dense \leo satellite systems. }
\red{The work of \cite{okati} uses stochastic geometry to investigate downlink interference and coverage probability in terms of \gsinr, assuming mutual satellite positions to be random processes.
	This work does not strictly capture how interference varies as a function of beams' steering directions and does not account for any explicit interference mitigation mechanisms.
}

\red{The authors in \cite{tonkin, dense_leo} conduct extensive assessments of the user link interference between \ngso systems in terms of throughput degradation, 
	applying various interference mitigation techniques such as look-aside and band-splitting. 
	The primary and secondary systems {were} assumed to randomly select their serving satellites from those above a certain elevation angle, 
	with the constraint that the secondary system's serving satellite was subject to a minimum specified angular separation %from the primary satellite-user link. 
	between its satellite-user link and the primary satellite-user link.
	Both studies \cite{tonkin, dense_leo} 
	acknowledge that the look-aside %method 
	helps decrease interference to some extent but has limited benefits for small constellations \cite{tonkin} and is inefficient under heterogeneous constellations, highly elliptical orbits (HEO) \cite{dense_leo}.
}%

\red{   
	While the studies of \cite{tonkin, dense_leo} are insightful, they do not provide a clear picture of the feasibility of spectrum sharing among non-cooperative mega-constellations of \leo satellites and do not explore more sophisticated approaches to secondary satellite selection beyond look-aside which guarantee some level of both downlink quality and protection.
	% It remains an open question
	Open questions remain regarding the number of secondary satellites which offer sufficiently low interference under techniques akin to look-aside, the bounds on interference between systems, {the prospects of secondary satellite selection to enable coexistence with incomplete knowledge of the primary system,} and how all of this varies as the satellites orbit the globe over time}.

Satellite system coexistence, channel estimation and prediction using machine learning for better performance. 

Satellite handovers - HE, MCT, .... 

Satellite channel estimation/prediction, .... 

Satellite coexistence.. .

The key is to predict the primary system - in particular, which primary satellite serves on the ground location of interest, so that the secondary system selects the secondary serving satellite on that ground area around the primary satellite and ground link to reduce interference. 

\subsection{Contribution}
}

%\subsection{Scope of the work}
%The completed feasibility study \cite{feasibility_journal} motivates ML approaches to satellite selections at the Kuiper system, which 1) maximizes the utilization (or \gsinr) of the Kuiper system and 2) meets the interference protection constraint. Considering the Kuiper system as an aggressor system responsible for coexistence with the incumbent system, our ML approaches are based on measurements taken by its ground users where no direct coordination or communication between these two systems exists.

%\vspace{-0.2cm}
\section{System Model}\label{sec:system_model}

In this section, we present the system model and key performance metrics necessary for analyzing the coexistence of two dense LEO satellite systems. 
We begin by defining the satellite systems' ground cell planning methodology over a wide geographical area. 
We then define the set of available satellites capable of serving these ground cells and establish performance metrics based on satellite-to-ground associations. This provides a structured framework for evaluating spectrum sharing and inter-system interference management in the coexistence scenario considered in this work.

% \begin{figure}
% 	\centering
% 	\includegraphics[width=\linewidth,height=0.25\textheight,keepaspectratio]{nn_fig/sat_system.pdf}
	
% 	\caption{System model of the proposed framework.}
%     \label{fig:sat_system}
% \end{figure}

\begin{figure}
    \vspace{-0.2cm}
	\centering    \includegraphics[width=\linewidth,height=0.18\textheight,keepaspectratio]{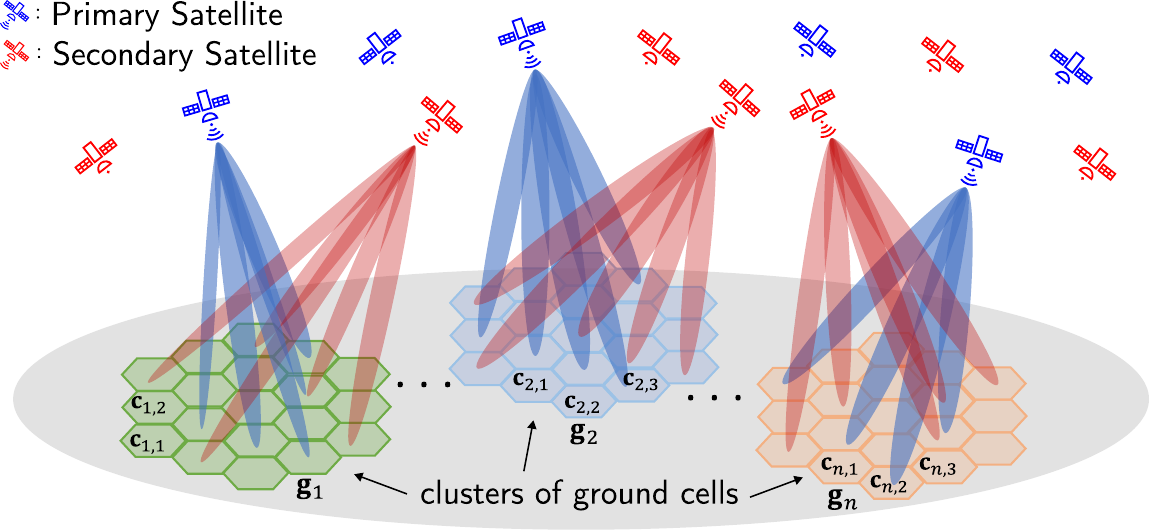}
	\caption{Primary and secondary satellite systems serve their ground users with multiple satellites, each forming $\numbeams$ spot beams. Cluster $\vg_n$ is served by a single satellite and consists of $\numc \gg \numbeams$ ground cells $\vc_{n, \ell}$, with each cell served by a single spot beam.}
    \label{fig:sat_system}
    \vspace{-0.4cm}
\end{figure}
\vspace{-0.25cm}
\subsection{Ground Cells and Clusters}
\begin{figure}[t]
    \vspace{-0.2cm}
	\centering	\includegraphics[width=\linewidth,height=0.21\textheight,keepaspectratio]{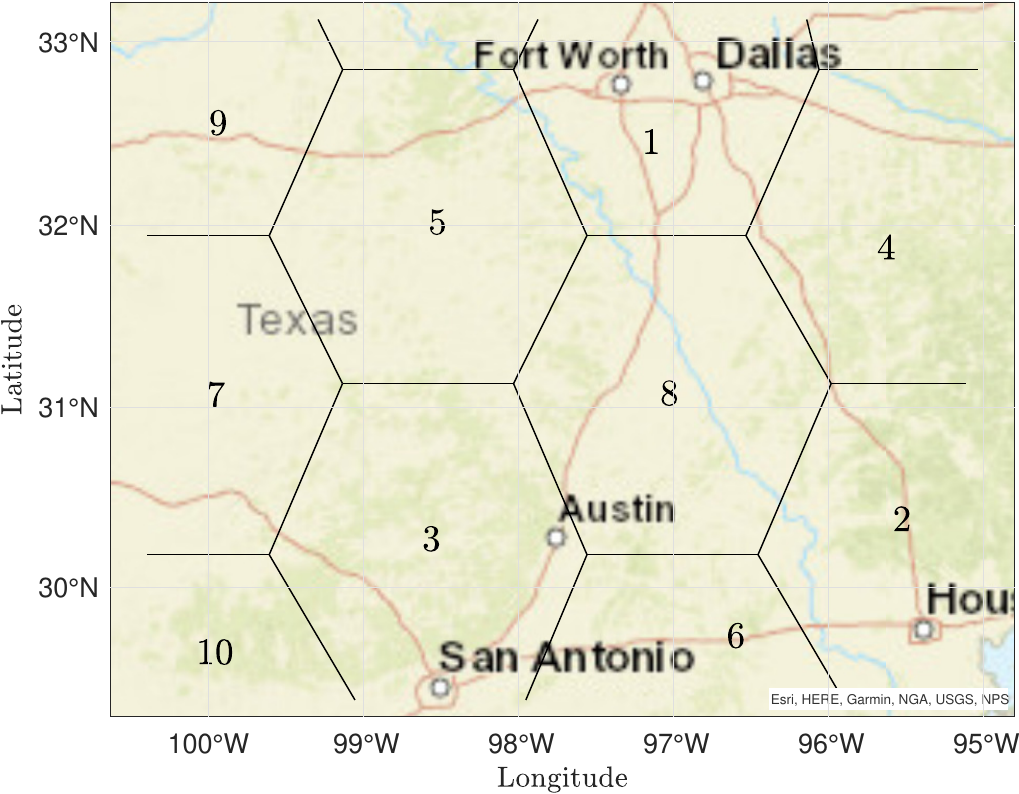}
	\caption{Geographical region considered in this work. Each hexagon represents a cluster with $\numc=127$ cells.}
    % In particular, the assigned number for each cluster indicates the priority of the cluster of the primary system. 
	\label{fig:geo_map}
    \vspace{-0.4cm}
\end{figure}

Satellites serve ground users by forming $\numbeams$ highly directional spot beams per satellite at once.
The area served by each spot beam is termed a \textit{cell}---typically spanning $300$--$500$~km\textsuperscript{2}~\cite{kuiper}---and the cells formed by a single satellite are tessellated on the ground, often in a hexagonal arrangement similar to those in terrestrial networks. %, but can be easily relaxed. 
With its multiple spot beams, each satellite is assumed responsible for serving a \textit{cluster} of $\numc$ cells during the time period between handovers. 
% across handover periods, e.g., $10$--$15$ sec. % over any given small time interval. 
In practical systems, it is often the case that $\numbeams \ll \numc$, with $\numbeams$ on the order of $16$ \cite{leo_comparison,num_spotbeams}, and thus each satellite must employ spatial, time, and/or frequency multiplexing~\cite{high_tput, kuiper} to provide adequate coverage over a wide area.
Each cluster of $\numc$ cells is fixed over time, though the satellite serving them will certainly change.
For simplicity of notation, let us assume the primary and secondary systems aim to serve their respective ground users distributed across overlapping cells and clusters, as illustrated in \figref{fig:sat_system}.
% For simplicity of notation, let us assume the primary and secondary systems aim to \red{independently} serve \red{their own ground users distributed in} the same cells and clusters, as illustrated in \figref{fig:sat_system}.
With this, let $\mathcal{G}$ be the set of $\numg$ clusters spanning some region of interest, defined as
\begin{align}
	\mathcal{G} = \braces{\vg_n: n={1, \dots, \numg}}, 
\end{align}
where $\vg_n$ denotes the $n$-th cluster of $\numc$ ground cells 
$\vg_n = \braces{\vc_{n,\ell}: \ell =1, \dots, \numcells}$. 
To provide some context, the geographical area of interest used in our numerical results is shown in \figref{fig:geo_map}, where each hexagon represents a cluster consisting of $\numc=127$ cells. 
This clustering coincides with coverage provided by practical multi-beam \leo satellites like Starlink~\cite{high_tput,spaceX_ss}. 
% This clustering is determined based on an analysis of the number of ground cells a satellite must serve, considering the coverage area and the available satellites in orbit \cite{high_tput}.
%\ipr{This clustering coincides with a general assumption of multi-beam satellite systems \cite{high_tput}. REMOVE?}
\comment{The number in each hexagon in the figure represents 
a form of prioritization of clusters in the order of satellite assignments.
In other words, clusters may have different priorities based on factors such as the number of subscribers. For example, given the limited number of satellites in orbit, denser areas may be prioritized to ensure higher service rates and better availability.}

\vspace{-0.25cm}
\subsection{Available Satellites and Ground Associations}

Let $\bar{\mathcal{P}}^t = \braces{\vp^t_i : i=1,\dots,\nump}$ denote the set of all $\nump$ satellites in the primary constellation at time $t$, with $\vp_i^t$ denoting the $i$-th primary satellite %\red{and its location} 
at time $t$. Similarly, let $\bar{\mathcal{S}}^t = \braces{\vs^t_m : m=1,\dots,\nums}$ be the set of all $\nums$ satellites in the secondary system's constellation at time $t$, with $\vs_m^t$ being the $m$-th secondary satellite %\red{and its location} 
at time $t$.
Out of all satellites in orbit, only those whose elevation angle---relative to a specified ground location---which exceeds the minimum elevation angle $\epsilon_{\mathrm{min}}$ %required by regulations 
can transmit signals toward %the vicinity of 
that ground location, according to regulations \cite{spaceX_ss, kuiper}. 
Thus, the \textit{overhead} sets of primary and secondary satellites capable of serving a particular ground cluster $\vg_n$ at time $t$ can be defined as
\begin{align}
    &\mathcal{P}^t_{n}  = \braces{\vp \in \overline{\setsatp}^t : \epsilon(\vg_n, \vp) \geq \epsilon_{\mathrm{min}}}, \\
    &\mathcal{S}^t_{n}  = \braces{\vs \in \overline{\setsats}^t : \epsilon(\vg_n, \vs) \geq \epsilon_{\mathrm{min}}}, \label{eq:min-elev}
\end{align}
where $\epsilon(\vg_n,\vx)$ denotes the minimum elevation of a satellite $\vx$ relative to the horizon at any location within ground cluster $\vg_n$.  
Building on this definition of overhead satellites, which is established with respect to a single cluster $\vg_n$, we  define the set of all satellites spanning clusters in $\mathcal{G}$ at time $t$ as
\begin{align}
    & \mathcal{P}^t_{\mathcal{G}} %\triangleq\mathcal{P}_{(n_1,  \cdots, n_\numg)}  
    = \setsatp^t_{1}  \cup\cdots  \cup  \setsatp^t_{\numg}, \label{eq:num_pavail} \\ %\quad n=1, \cdots, \numg,
    & \mathcal{S}^t_{\mathcal{G}} %\triangleq\mathcal{S}_{(n_1,  \cdots, n_\numg)}  
    = \setsats^t_{1}  \cup\cdots  \cup  \setsats^t_{\numg}.
    \label{eq:num_savail}
\end{align}
%\blue{where $M^t = |\mathcal{S}^t_{\mathcal{G}}|$. }
We are particularly interested in how the secondary system satellites are associated to serve each cluster. 
Let us therefore define $x^t_{m,n}$ as a binary satellite-to-cluster association variable 
\begin{align}
    x^t_{m,n} = 
    \begin{cases}
    1, & \parbox[t]{5.5cm}{if the $m$-th secondary satellite serves cells in cluster $\vg_n$ at time $t$} \\
	% 1, & {\text{if } \text{the } m\text{-th secondary satellite serves cluster }\vg_n\text{at time } $t$}\\% \text{ at time }t,\\
		0, & \text{otherwise}.
        \label{eq:x}
	\end{cases}
\end{align}
Let us denote by $\bX^t \in \braces{0,1}^{\nums\times\numg}$ the secondary system association matrix at time $t$ containing all binary association variables $x_{m,n}^t$.
% Considering the limited number of physical spot beams, 
Considering each satellite has a limited capacity and limited number of spot beams \cite{leo_comparison,num_spotbeams}, we assume that clusters are sized such that each satellite may only serve one cluster at any given time $t$, leading to the constraint
% We have the condition on the number of clusters a satellite may serve at time $t$
\begin{align}
    \sum_{n=1}^{\numg} x^t_{m,n} \leq 1 \ \forall  \ m = 1, 2, \dots, \nums, 
\end{align}
% where $M\geq 1$. 
% This work assumes $M=1$ as the number of satellites $\nums$ of considered constellations is far fewer than the number of ground cells on the surface of the Earth. 
% Also,since a single cluster of cells may only be served by one satellite, we have the condition
It is also reasonable to assume that, at a given frequency, a single cluster may only be served by one satellite at any given time to avoid intra-system interference, meaning
\begin{align}
    \sum_{m=1}^{\nums} x^t_{m,n} \leq 1 \  \forall \ n = 1, 2, \dots, \numg.
\end{align}
Having assumed a single satellite may not necessarily serve all cells in a cluster at a given time due to a limited number of spot beams, we also define the binary satellite-to-cell association variable of the secondary system at time $t$ within a cluster as
\begin{align}
    y^t_{m,n,\ell} = 
    \begin{cases}
        1, & \parbox[t]{5.5cm}{if the $m$-th secondary satellite serves the $\ell$-th cell $\vc_{n,\ell}$ in cluster $n$ at time $t$} \\
		% 1, & \text{if } \text{the } m\text{-th secondary satellite serves cell } \vc_{n,\ell},\\
		0, & \text{otherwise}. \label{eq:y}
	\end{cases} 
\end{align}
% where $\vc_{n,\ell}$ is the $\ell$-th cell of cluster $\vg_n$. 
The number of cells served by a satellite at once within a single cluster is bounded by its number of spot beams, i.e.,  %  $K$. 
\begin{align}
    \sum_{\ell=1}^{\numc} y^t_{m,n,\ell} \leq \numbeams \ \forall \ m,n,t.
    \vspace{-0.3cm}
\end{align}

%\vspace{-0.25cm}
\subsection{Key Performance Metrics}
Let us define the transmit antenna gain of primary satellite $\vp$ in the direction of a primary user $\sfu$ as $\Gtx(\sfu,\vp,\vc)$ when $\vp$ steers its beam toward $\vc$. 
Similarly, let $\Grx(\sfu,\vp)$ denote the receive antenna gain of the primary user $\sfu$ in the direction of the primary satellite $\vp$. %  when $\vp$ steers its beam toward cell $\vc$. 
Path loss between a user $\sfu$ and satellite $\vp$ is denoted by $L(\sfu,\vp)$ and modeled as \cite{3gpp38811}
\begin{align}
    \todB{L(\sfu, \vp)} = 32.45 + 20\,\logten{f} + 20\,\logten{d(\sfu, \vp)},
\end{align}
where $f$ is the carrier frequency and $d(\sfu, \vp)$ is the path distance between $\sfu$ and $\vp$.
Based on these definitions, the \gsnr at a primary ground user $\sfu$ in cell $\vc$ when receiving from its serving satellite $\vp$ is given by
\begin{align}
    \msnr(\sfu,\vp, \vc) = \frac{\powertx(\vp) \ \Gtx(\sfu,\vp, \vc) \ \Grx(\sfu,\vp)}{\powernoise(\sfu) \ L(\sfu,\vp)}, 
    \label{eq:u_snr}
\end{align}
where $\powertx(\vp)$ represents the transmit power of satellite $\vp$, and $\powernoise(\sfu)$ denotes the noise power at user $\sfu$.

In this work, we are interested in the downlink interference inflicted onto a primary user $\sfu$ by secondary satellites. 
Let us thus suppose a secondary satellite $\vs$ steers its beam toward cell $\vc$ and, in doing so, inflicts interference onto primary ground user $\sfu$ being served by a satellite $\vp$. 
Then, the resulting \inr at the primary ground user $\sfu$ can be written as 
\begin{align} \label{eq:inr-p}
\minr(\sfu,\vp;\vs, \vc) = \frac{\powertx(\vs) \ \Gtx(\sfu;\vs,\vc) \ \Grx(\sfu,\vp;\vs,\vc)}{\powernoise(\sfu) \ L(\sfu,\vs)}.
\end{align}
Here, we have slightly extended the notation of $\Gtx(\cdot)$ and 
$\Grx(\cdot)$ to make it clear that $\Gtx(\sfu;\vs,\vc)$ represents the transmit gain of satellite $\vs$ in the direction of the primary user $\sfu$ when $\vs$ serves its ground cell $\vc$.
Similarly, $\Grx(\sfu,\vp;\vs,\vc)$ represents the receive gain in the direction of the secondary satellite $\vs$ when the %ground 
user $\sfu$ steers its beam toward its serving satellite $\vp$.  

Having considered multiple satellites serving multiple ground cells, we are concerned with the \textit{aggregate} interference inflicted onto a given primary user by all active spot beams in the secondary system. %  (and vice versa). 
Let us define the collective interference inflicted by a single secondary satellite $\vs_m$ onto a primary user $\sfu$ at time $t$ as
\begin{align} 
    \minr^t(\sfu, \vp;\vs^t_m, \vg_n) 
    \triangleq
    \sum_{\ell=1}^{\numc} \minr^t(\sfu, \vp; \vs^t_m, \vc_{n, \ell}) \cdot y^t_{m,n, \ell}, %\label{eq:pinr_def} 
\end{align}
where the user has steered its beam toward satellite $\vp$. 

Recall, the binary association variable $x^t_{m,n}$ indicates whether secondary satellite $\vs_m$ actively serves cluster $\vg_n$ at time $t$, thereby contributing interference to the primary user $\sfu$. 
The aggregate interference inflicted upon primary user $\sfu$ by all secondary satellites at time $t$ can thus be stated as
\begin{align} 
    \minr^t(\sfu, \vp) \triangleq \sum_{m=1}^{\nums}\sum_{n=1}^\numg \minr^t(\sfu, \vp; \vs^t_m, \vg_n) \cdot x^t_{m,n}. \label{eq:pinr_def} 
\end{align}
%Here, we have made the minor simplification that satellites outside of the overhead set $\mathcal{S}^t_{\mathcal{G}}$ inflict negligible interference. 
The \gsinr of a primary user $\sfu$ in cell $\vc$ when served by satellite $\vp$ at time $t$ is then  %written as
\begin{align}
    \msinr^t(\sfu,\vp, \vc) =  \frac{\msnr^t(\sfu, \vp, \vc)}{1 + \minr^t(\sfu, \vp)},
    \label{eq:psinr}
\end{align}
where intra-system interference inflicted by the primary system onto its own users has been ignored for simplicity.
% but could be straightforwardly absorbed into the effective noise power of its users.
In a similar manner, the \gsinr of a secondary user $\sfv$ served by secondary satellite $\vs$ at time $t$ is of the form 
\begin{align}
    \msinr^t(\sfv,\vs, \vc) =  \frac{\msnr^t(\sfv, \vs, \vc)}{1 + \minr^t(\sfv, \vs)},
    \label{eq:ssinr}
\end{align}
where $\minr^t(\sfv, \vs)$ is analogously the aggregate interference inflicted upon $\sfv$ by the primary system.
\section{Optimizing Secondary Satellite\\Selection for Coexistence} \label{sec:sat_selection}

In our prior work \cite{feasibility_journal}, we showed that there is promise in enabling the \leo coexistence paradigm laid forth through the strategic selection of secondary serving satellites---courtesy of the spatial diversity across primary and secondary satellite constellations at virtually any given time.
Creating practical mechanisms to perform such a selection on a network scale remains an open problem, however, which amounts to optimizing the secondary system satellite-cluster association matrix $\bX^t$ over time $t$ to maximize a desired objective while abiding by certain constraints. 
More specifically, in selecting which secondary satellites serve its ground users, we aim to maximize the downlink performance of the secondary satellite system while guaranteeing primary ground users are protected from a certain level of interference inflicted by secondary satellites.
% In this pursuit, we first formulate our problem and then
% That is, we aim to determine the near-optimal 

% In this section, we investigate mechanisms for the secondary system to select satellites that enable coexistence with the primary system while maximizing the capacity of the secondary system. 
\vspace{-0.2cm}
\subsection{Interference Protection Constraint}

\setcounter{equation}{21}
\begin{figure*}[t]
    \vspace{-0.3cm}
	% ensure that we have normalsize text
	\normalsize
	% Store the current equation number.
	%	\setcounter{mytempeqncnt}{\value{equation}}
	%\setcounter{equation}{4}
	% Set the equation number to one less than the one
	% desired for the first equation here.
	% The value here will have to changed if equations
	% are added or removed prior to the place these
	% equations are referenced in the main text.
	%	\setcounter{equation}{4}
	\begin{align}
    \max_{\sfu\in\mathcal{U}} \, 
    \underset{\tau \in [t-\Tw, t+\Th)}{\E}\brackets{\minr^\tau(\sfu, \vp^\tau_\sfu)} 
    % &= 
    % \max_{\sfu\in\mathcal{U}} \, 
    % \frac{1}{T_\mathrm{w}+T_\mathrm{h}} \brackets{\sum_{\tau=t-T_\mathrm{w}}^{t-1} \minr^{\tau}(\sfu, \vp^{\tau}_\sfu) + \sum_{\tau=t}^{t+T_\mathrm{h}-1} \minr^{\tau}(\sfu, \vp^{\tau}_\sfu)} \\
    &\leq
    \frac{1}{T_\mathrm{w}+T_\mathrm{h}} \brackets{
    \max_{\sfu\in\mathcal{U}} \, 
    \sum_{\tau=t-T_\mathrm{w}}^{t-1} \minr^{\tau}(\sfu, \vp^{\tau}_\sfu) + 
    \max_{\sfu\in\mathcal{U}} \, 
    \sum_{\tau=t}^{t+T_\mathrm{h}-1} \minr^{\tau}(\sfu, \vp^{\tau}_\sfu)
    }\label{eq:constraint-split}
    \end{align}
	% Restore the current equation number.
	%	\setcounter{equation}{\value{mytempeqncnt}}
	%	\setcounter{equation}{5}
	% IEEE uses as a separator
	% \hrulefill
	% The spacer can be tweaked to stop underfull vboxes.
	% \vspace*{4pt}
    \vspace{-0.3cm}
\end{figure*}
\begin{figure*}[t]
        \vspace{-0.5cm}%
	% ensure that we have normalsize text
	\normalsize
	% Store the current equation number.
	%	\setcounter{mytempeqncnt}{\value{equation}}
	%\setcounter{equation}{4}
	% Set the equation number to one less than the one
	% desired for the first equation here.
	% The value here will have to changed if equations
	% are added or removed prior to the place these
	% equations are referenced in the main text.
	%	\setcounter{equation}{4}
	\begin{align}
    \max_{\sfu\in\mathcal{U}} \, 
    \frac{1}{\Th} \sum_{\tau=t}^{t+T_\mathrm{h}-1} \minr^{\tau}(\sfu, \vp^{\tau}_\sfu) 
    \leq 
    \underbrace{%
    \frac{1}{\Th}
    \brackets{%
    \parens{\Tw + \Th} \cdot \tminrth - \max_{\sfu\in\mathcal{U}}
    \sum_{\tau=t-T_\mathrm{w}}^{t-1} \minr^{\tau}(\sfu, \vp^{\tau}_\sfu)%
    }%
    }_{\triangleq \, \tminrth^t} \label{eq:pinr_const_max_new}
    % &=
    % \frac{1}{N_\mathrm{w}+N_\mathrm{h}+1}\brackets{\sum_{n=-N_\mathrm{w}}^{N_\mathrm{h}} \minr^{t+nT}(\sfu, \vp^{t+nT}_\sfu;\vs_m,\vg_n) + \sum_{n=-N_\mathrm{w}}^{N_\mathrm{h}} \minr^{t+nT}(\sfu, \vp^{t+nT}_\sfu;\vs_m,\vg_n)},
    %\label{eq:pinr_exp}
\end{align}
	% Restore the current equation number.
	%	\setcounter{equation}{\value{mytempeqncnt}}
	%	\setcounter{equation}{5}
	% IEEE uses as a separator
	% \hrulefill
	% The spacer can be tweaked to stop underfull vboxes.
	% \vspace*{4pt}
    \vspace{-0.3cm}%
\end{figure*}
\setcounter{mytempeqncnt}{\value{equation}}%

\begin{figure*}[t]
    \vspace{-0.5cm}%
	% ensure that we have normalsize text
	\normalsize%
	% Store the current equation number.
	%	\setcounter{mytempeqncnt}{\value{equation}}
	%\setcounter{equation}{4}
	% Set the equation number to one less than the one
	% desired for the first equation here.
	% The value here will have to changed if equations
	% are added or removed prior to the place these
	% equations are referenced in the main text.
	%	\setcounter{equation}{4}
    \begin{align}
        \frac{1}{\Th}
        \sum_{\tau=t}^{t+T_\mathrm{h}-1} \minr^{\tau}(\sfu, \vp^{\tau}_\sfu) 
        % &= 
        % \frac{1}{\Th}
        % \sum_{\tau=t}^{t+T_\mathrm{h}-1} 
        % \sum_{m=1}^{\nums}\sum_{n=1}^\numg \minr^\tau(\sfu, \vp^\tau_\sfu; \vs_m, \vg_n) \cdot x^\tau_{m,n} \\
        % &= 
        % \sum_{m=1}^{\nums}\sum_{n=1}^\numg 
        % \frac{1}{\Th}
        % \sum_{\tau=t}^{t+T_\mathrm{h}-1}
        % \minr^\tau(\sfu, \vp^\tau_\sfu; \vs_m, \vg_n) \cdot x^\tau_{m,n} \\
         = 
        \sum_{m=1}^{\nums}\sum_{n=1}^\numg 
        \underbrace{\frac{1}{\Th} \sum_{\tau=t}^{t+T_\mathrm{h}-1}
        \minr^\tau(\sfu, \vp^\tau_\sfu; \vs_m, \vg_n)}_{\triangleq \, \tminrTht(\sfu, \vp_\sfu; \vs_m, \vg_n)}\, x^t_{m,n}. %\minr^{t}_{\sum_\Th}(\sfu, \vp_\sfu; \vs_m, \vg_n)}.
        \label{eq:avg_inr_Th}
    \end{align}

	% Restore the current equation number.
	%	\setcounter{equation}{\value{mytempeqncnt}}
	%	\setcounter{equation}{5}
	% IEEE uses as a separator
	\hrulefill
	% The spacer can be tweaked to stop underfull vboxes.
	%\vspace*{4pt}
    \vspace{-0.3cm}
\end{figure*}%
\setcounter{equation}{17}%

% \begin{figure}[t]
% 	\centering
%     \vspace{-0.3cm}	\includegraphics[width=\linewidth,height=0.23\textheight,keepaspectratio]{nn_fig/leo_coex_type_07.pdf}
%     \caption{Combining a time-average constraint $\tminrth$ and an absolute constraint $\minrthmax$ provides flexibility in secondary satellite selection while guaranteeing a certain level of protection to primary users. %In turn, this can yield fruitful coexistence of the primary and secondary \leo constellations.
%     }
%     \label{fig:pinr_const}
%     \vspace{-0.4cm}
% \end{figure}

The key condition for the secondary system to coexist with the primary system is that it must not inflict prohibitive interference on the primary system. 
Defining an interference protection constraint is therefore important in the context of this work, especially since the precise definition of prohibitive interference remains fluid, especially on a global scale \cite{ntia_ipc, 47cfr25_261}. 
Recall, the aggregate interference inflicted by the secondary system upon a given ground user $\sfu$ at time $t$ is denoted by $\minr^t(\sfu, \vp_\sfu)$ and is calculated by summing the interference inflicted by all active spot beams of satellites, as defined in \eqref{eq:pinr_def}. 
Perhaps the simplest yet strictest way to define an interference protection constraint is therefore to require that the aggregate interference be below a specified threshold $\minrth$ at all primary users and at all times, i.e., 
\begin{align}
    \minr^t(\sfu, \vp^t_\sfu) \leq \minrth \ \forall \ \sfu \in \mathcal{U}, \ t\in[0,\infty),
    \label{eq:strict_ipr}
\end{align}
where $\vp^t_\sfu$ is the serving satellite of $\sfu$ at time $t$ and $\mathcal{U}$ is the set of all primary ground users to protect.
While this strict constraint is intuitive and certainly very protective, we have found it to be too restrictive for realistic thresholds $\minrth$, often leading to poor utilization of the secondary system, as we will show in numerical results.
Furthermore, when employing this strict constraint, we found that interference levels fall well below $\minrth$ for most users the majority of the time.
% In other words, setting a threshold of $\minrth$ often yields an interference distribution that falls orders of magnitude below $\minrth$, except for its upper tail.
This motivates us to consider more flexible still well-defined interference protection constraints.

In this pursuit, we formalize three proposed principles.
First, since handover in practical \leo systems usually happens on time scales of seconds or tens of seconds \cite{starlink_measure, starlink_measure2}, protection should be proactive in the sense that it should be guaranteed throughout the time period between handovers.
Second, given the complexity and time-varying nature of this coexistence paradigm, infrequent, short-lived spikes in interference at a given user are extremely difficult to completely eliminate without frequent service interruptions and should thus be forgiven to a certain extent, to facilitate plausible coexistence. 
Third, interference should nonetheless never exceed a specified threshold at any given primary user.
% Crafting an interference protection constraint with these three principles in mind will ensure protection is maintained between handovers while tolerating what would otherwise be infrequent violations of protection.

To properly formulate a constraint with these principles in mind, suppose we are interested in performing secondary satellite-cluster association at time $t$, i.e., we seek to optimize $\bX^{t}$.
Let us denote by $T_\mathrm{h} > 0$ the time between secondary system handovers, and assume that secondary association is fixed between handover instances, after which a new association will be made.
Put simply, if a handover is performed at time $t$, then $\bX^{t} = \bX^{\tau} \ \forall \ \tau \in [t,t+T_\mathrm{h})$.
It is important to keep in mind, however, that interference levels will nonetheless fluctuate throughout the handover period $[t,t+T_\mathrm{h})$, since both the secondary and primary satellites will move along their orbits.
Now, if interference had been relatively high leading up to handover time $t$, it is sensible to be less tolerable of interference during the upcoming handover period $[t,t+T_\mathrm{h})$, and vice versa.
% Similarly, if interference had been relatively low up to time $t$, it is perhaps reasonable to tolerate higher interference.
To capture this, 
% let $T_\mathrm{w} \geq 0$ be the duration of a historical time window considered when constraining the association $\bX^{t}$.
let us formulate the following constraint on the time-averaged interference across the time horizon $[t-T_\mathrm{w},t+T_\mathrm{h})$: %  can thus be stated as
\begin{align}
    \underset{\tau \in [t-\Tw, t+\Th)} {\E}\brackets{\minr^\tau(\sfu, \vp^\tau_\sfu)} 
    \leq \tminrth \ \forall \ \sfu \in \mathcal{U},
    % \underset{\tau \in t + \mathcal{N}}{\E}
    % \brackets{ \minr^\tau(\sfu, \vp_\sfu)} \leq \minrth, \quad\forall t, 
    \label{eq:pinr_const}
\end{align}
%for all $t \in \braces{k\Th: k=0, 1, \dots}$ 
for all handover times $t$, 
where $\Tw \geq 0$ is the duration of the past time window considered when averaging interference. % duration of the historical time window. 
\comment{Notice that, since \eqref{eq:pinr_const} must be satisfied for all primary users $\sfu \in \mathcal{U}$, it can be written equivalently as 
\begin{align}
    \max_{\sfu\in\mathcal{U}} \, \underset{\tau \in [t-\Tw, t+\Th)}{\E}\brackets{\minr^\tau(\sfu, \vp^\tau_\sfu)} 
    \leq \tminrth, % \ \forall \ t.
    \label{eq:pinr_const_max}
\end{align}
for all handover times $t$. 
}

Ultimately, our satellite-cluster association problem amounts to finding the optimal sequence of associations at each handover instance across all time, i.e., $\braces{\bX^t : t = k\Th, k=0,1,\dots}$. Since our time-averaged constraint \eqref{eq:pinr_const} spans more than one handover period when $\Tw > 0$, the association at time $t$ impacts the association at time $t+\Th$ and thus finding an optimal sequence of associations $\braces{\bX^t}$ is a joint optimization problem. % (whose objective is yet to be defined)
% This is a challenging, non-convex task
The challenging combinatorial nature of this problem motivates us to tackle it in a sequential manner by solving for $\bX^{t-\Th}$ at time $t-\Th$, then for $\bX^{t}$ at time $t$, and so on.
% then for $\bX^{t+\Th}$ at time $t+\Th$, and so on.
In doing so, the interference inflicted up to time $t$ becomes fixed when solving for $\bX^{t}$ and makes this problem more manageable, as will become clear next.
% % i.e., the interference during the time period $[t-\Tw,t)$, 

%\setcounter{mytempeqncnt}{\value{equation}}
%\ipr{Many things to fix here.}
Henceforth, let us discretize time and treat $t$ as a time slot index and $T_\mathrm{w}$ and $T_\mathrm{h}$ have being in units of time slots. 
Then, we can write the left-hand side of \eqref{eq:pinr_const} as the summation
\begin{align}
   &\! \underset{\tau \in [t-\Tw, t+\Th)}
    {\E}\brackets{\minr^\tau(\sfu, \vp^\tau_\sfu)} \nonumber\\
    % &= 
    % \frac{1}{T_\mathrm{w}+T_\mathrm{h}} \int_{t-T_\mathrm{w}}^{t+T_\mathrm{h}}     \minr^\tau(\sfu, \vp^\tau_\sfu) \, \mathrm{d}\tau \\
    % &\approx
    % &=
    % \frac{1}{T_\mathrm{w}+T_\mathrm{h}}\sum_{\tau=t-T_\mathrm{w}}^{t+T_\mathrm{h}-1} \minr^{\tau}(\sfu, \vp^{\tau}_\sfu), \\
    &\!\!=
    \frac{1}{T_\mathrm{w}+T_\mathrm{h}} \!\brackets{\sum_{\tau=t-T_\mathrm{w}}^{t-1} \!\minr^{\tau}(\sfu, \vp^{\tau}_\sfu)\! + \!\sum_{\tau=t}^{t+T_\mathrm{h}-1} \!\minr^{\tau}(\sfu, \vp^{\tau}_\sfu)},\!
    \label{eq:pinr_exp}
\end{align}
where we have split the sum into the interference inflicted during $[t-\Tw,t)$ and that during $[t,t+\Th)$.
Suppose we aim to find an association $\bX^t$ at time $t$ that satisfies constraint \eqref{eq:pinr_const}, and solely for the sake of discussion, let us suppose $\Tw \leq \Th$.
Given some association $\bX^{t-\Th}$ made at time $t-\Th$ and used throughout the handover period $[t-\Th,t)$, satisfying constraint \eqref{eq:pinr_const} depends only on the second summation in \eqref{eq:pinr_exp} spanning $[t,t+\Th)$, since the first summation depends only on $\bX^{t-\Th}$.

Since \eqref{eq:pinr_const} must be satisfied for all primary users $\sfu \in \mathcal{U}$, it can be written equivalently as 
\begin{align}
    \max_{\sfu\in\mathcal{U}} \, \underset{\tau \in [t-\Tw, t+\Th)}{\E}\brackets{\minr^\tau(\sfu, \vp^\tau_\sfu)} 
    \leq \tminrth, % \ \forall \ t.
    \label{eq:pinr_const_max}
\end{align}
for all handover times $t$.
Plugging \eqref{eq:pinr_exp} into \eqref{eq:pinr_const_max} and distributing the maximum across the two summations allows us to arrive at inequality \eqref{eq:constraint-split}.
% and thus \eqref{eq:pinr_exp} can be written as \eqref{eq:constraint-split}.}
%We are thus motivated to split the left-hand side of constraint \eqref{eq:pinr_const_max} into two by the inequality as in \eqref{eq:constraint-split}. 
Note that by ensuring the right-hand side of \eqref{eq:constraint-split} is less than the threshold $\tminrth$, the left-hand side is guaranteed to also be below $\tminrth$. %  that threshold.
% Substituting the left-hand side of \eqref{eq:pinr_const_max} by the right-hand side of \eqref{eq:constraint-split}, we get
Substituting the right-hand side of \eqref{eq:constraint-split} into the left-hand side of \eqref{eq:pinr_const_max} 
% we get
% \begin{align}
% \frac{1}{T_\mathrm{w}+T_\mathrm{h}} \brackets{
% \max_{\sfu\in\mathcal{U}} \, 
% \sum_{\tau=t-T_\mathrm{w}}^{t-1} \minr^{\tau}(\sfu, \vp^{\tau}_\sfu) + 
% \max_{\sfu\in\mathcal{U}} \, 
% \sum_{\tau=t}^{t+T_\mathrm{h}-1} \minr^{\tau}(\sfu, \vp^{\tau}_\sfu)}
% \leq 
% \tminrth
% \ \forall \ t.
% \end{align}
and rearranging terms, we arrive at \eqref{eq:pinr_const_max_new}, 
% \begin{align}
%     \max_{\sfu\in\mathcal{U}} \, 
%     \frac{1}{\Th} \sum_{\tau=t}^{t+T_\mathrm{h}-1} \minr^{\tau}(\sfu, \vp^{\tau}_\sfu) 
%     \leq 
%     \underbrace{%
%     \frac{1}{\Th}
%     \brackets{%
%     \parens{\Tw + \Th} \cdot \tminrth - \max_{\sfu\in\mathcal{U}}
%     \sum_{\tau=t-T_\mathrm{w}}^{t-1} \minr^{\tau}(\sfu, \vp^{\tau}_\sfu)%
%     }%
%     }_{\triangleq \, \tminrth^t}%
%     \ \forall \ t, \label{eq:pinr_const_max_new}
%     % &=
%     % \frac{1}{N_\mathrm{w}+N_\mathrm{h}+1}\brackets{\sum_{n=-N_\mathrm{w}}^{N_\mathrm{h}} \minr^{t+nT}(\sfu, \vp^{t+nT}_\sfu;\vs_m,\vg_n) + \sum_{n=-N_\mathrm{w}}^{N_\mathrm{h}} \minr^{t+nT}(\sfu, \vp^{t+nT}_\sfu;\vs_m,\vg_n)},
%     %\label{eq:pinr_exp}
% \end{align}
%whose left-hand side depends on the association $\bX^t$ while the right-hand side is instead determined by prior associations. 
whose right-hand side is deterministic at time $t$, assuming prior associations have been made.
% In other words, 
Put simply, only the left-hand side of \eqref{eq:pinr_const_max_new} depends on $\bX^t$.
For this reason, we denote by $\tminrth^t$ the time-averaged interference threshold imposed on the association $\bX^t$ made at time $t$. %  throughout the handover period $[t,t+\Th)$.

\begin{figure}[t]
	\centering
    \vspace{-0.2cm}
    \includegraphics[width=\linewidth,height=0.23\textheight,keepaspectratio]{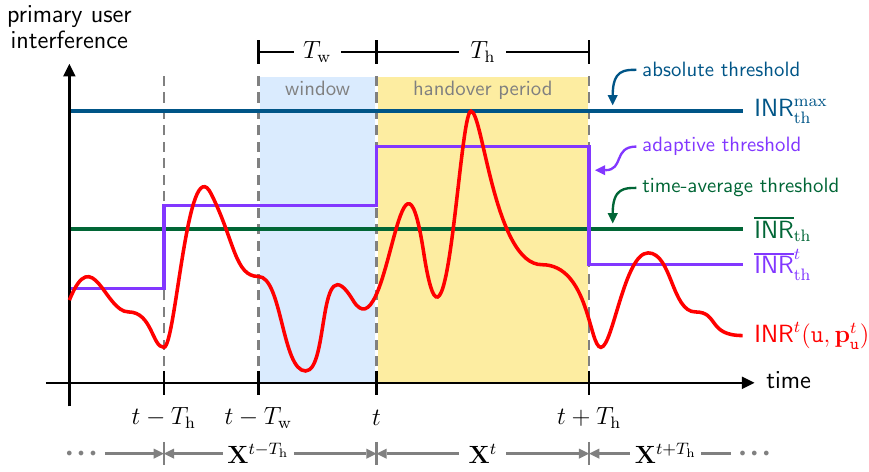}
    \caption{Combining a time-average constraint $\tminrth$ and an absolute constraint $\minrthmax$ provides flexibility in secondary satellite selection while guaranteeing a certain level of protection to primary users. %In turn, this can yield fruitful coexistence of the primary and secondary \leo constellations.
    }
    \label{fig:pinr_const}
    \vspace{-0.4cm}
\end{figure}

\setcounter{equation}{24}%
Further, leveraging the fact that $x^\tau_{m,n} = x^t_{m,n}$ for all $\tau \in [t,t+T_\mathrm{h})$, we arrive at \eqref{eq:avg_inr_Th}, 
% Employing this definition of $\tminrTht(\sfu, \vp_\sfu; \vs_m, \vg_n)$ %$\minr^{t}_{\sum_\Th}(\sfu, \vp_\sfu; \vs_m, \vg_n)$
%This allows 
allowing us to simplify \eqref{eq:pinr_const_max_new} as
\begin{align}
    \max_{\sfu\in\mathcal{U}} \, 
    \sum_{m=1}^{\nums}\sum_{n=1}^\numg 
    \tminrTht(\sfu, \vp_\sfu; \vs_m, \vg_n)  \, x^t_{m,n}
   % \mathbbm{1}\brackets{\vs_m\in\setsats^t_n} \nonumber\\
    \leq 
    \tminrth^t, %
    % \ \forall \ \sfu, \ t, 
    \label{eq:pinr_const_max_newer}
\end{align}
where $\tminrTht(\sfu, \vp_\sfu; \vs_m, \vg_n)$ is defined in \eqref{eq:avg_inr_Th}.
%$\vs_m\in \mathcal{S}^t_n$
For any associations made up until time $t$, satisfying \eqref{eq:pinr_const_max_newer} when optimizing the association $\bX^t$ will ensure that the time-averaged interference inflicted upon any primary user $\sfu \in \mathcal{U}$ over the time horizon $[t-\Tw,t+\Th)$ will be less than a specified threshold $\tminrth$.
Since this time-averaged constraint \eqref{eq:pinr_const_max_newer} imposes no bounds on interference at any given instant, an absolute constraint can be employed in tandem with \eqref{eq:pinr_const_max_newer} as
\begin{align}
\max_{\substack{\sfu \in \mathcal{U}\\\tau \in \mathcal{T}}}\sum_{m=1}^{\nums} \sum_{n=1}^{\numg} \minr^\tau(\sfu, \vp^t_\sfu; \vs_m, \vg_n)x_{m,n}^t \ \leq \ \minrth^{\mathrm{max}},
%\max_{\sfu \in \mathcal{U},\tau\in \mathcal{T}}
% \sum_{m=1}^\nums \sum_{n=1}^\numg \minr^\tau(\sfu, \vp^t_\sfu; \vs, \vg_n) \, \ x_{m,n}^t \leq \minrth^{\mathrm{max}} \ \forall \  \tau \in \mathcal{T}, % , . 
\end{align} % \sfu\in\mathcal{U}, \,
for some threshold $\minrth^{\mathrm{max}} \geq \tminrth$, where $\mathcal{T}= [t, t+\Th)$ denotes the handover period.
This ensures that the interference inflicted upon each primary user does not exceed some desired threshold at any given time during the entire handover period $\mathcal{T}$.
\figref{fig:pinr_const} illustrates how our proposed time-averaged interference constraint and absolute constraint ensure a desired level of protection is maintained across mutliple handover periods. 
% \figref{fig:pinr_const} illustrates how the instantaneous $\minr^t(\sfu, \vp^t_\sfu)$ is regulated by the time-average constraint %$\tminrth$ 
% and absolute threshold %$\minrthmax$ 
% over multiple handover periods, as a function of $\Tw$, $\Th$, and $\ttminrth$.

\setcounter{equation}{28}
\begin{figure*}[t]
    \vspace{-0.3cm}
	% ensure that we have normalsize text
	\normalsize
        % \small
	% Store the current equation number.
	%	\setcounter{mytempeqncnt}{\value{equation}}
	%\setcounter{equation}{4}
	% Set the equation number to one less than the one
	% desired for the first equation here.
	% The value here will have to changed if equations
	% are added or removed prior to the place these
	% equations are referenced in the main text.
	%	\setcounter{equation}{4}
    \begin{multline}
    g(\bX^t, \lambda, \mu, \boldsymbol\nu)
    =
    \sum_{n=1}^{\numg} {\sum_{m=1}^{\nums} 
    \parens{c^t_{m,n} 
    - \lambda \cdot \max_{\sfu\in\mathcal{U}} \, \tminrTht(\sfu, \vp_\sfu; \vs_m, \vg_n)  
    -\mu \cdot \max_{\sfu\in\mathcal{U}, \tau \in \mathcal{T}} \minr^\tau(\sfu, \vp_\sfu; \vs_m, \vg_n)
    - \nu_m
    } \cdot x^t_{m,n}}\\%_{\triangleq \, g_n(\lambda, \mu, \boldsymbol\nu)} \\
    + \lambda \cdot \ttminrth
    +\mu\cdot \minrth^{\mathrm{max}}
    + \sum_{m=1}^{\nums}\nu_m  \label{eq:ld}
    \end{multline}

	% Restore the current equation number.
	%	\setcounter{equation}{\value{mytempeqncnt}}
	%	\setcounter{equation}{5}
	% IEEE uses as a separator
	%\hrulefill
	% The spacer can be tweaked to stop underfull vboxes.
	%\vspace*{4pt}
    \vspace{-0.2cm}
\end{figure*}

\setcounter{equation}{31}
\begin{figure*}[t]
    \vspace{-0.6cm}
	% ensure that we have normalsize text
	\normalsize
	% Store the current equation number.
	%	\setcounter{mytempeqncnt}{\value{equation}}
	%\setcounter{equation}{4}
	% Set the equation number to one less than the one
	% desired for the first equation here.
	% The value here will have to changed if equations
	% are added or removed prior to the place these
	% equations are referenced in the main text.
	%	\setcounter{equation}{4}
	\begin{align}
	 \max_{\bX^t} &\sum_{m=1}^{N_s}\braces{ c_{m,n}^t-\lambda \cdot \max_{\sfu\in\mathcal{U}} \, \tminrTht(\sfu, \vp_\sfu; \vs_m, \vg_n) -\mu \cdot \max_{\sfu\in\mathcal{U}, \tau \in \mathcal{T}} \minr^\tau(\sfu, \vp_\sfu; \vs_m, \vg_n) - \nu_m}\cdot x_{m,n}^t \ \forall \ n \label{eq:sp_1}
		% \st & x_{m,n} \in \braces{0,1},\,\,\forall m, \\
		%&\sum_{m=1}^{N_s} x_{m,n} \leq 1. \label{eq:sp_3}
	\end{align}
	
	% Restore the current equation number.
	%	\setcounter{equation}{\value{mytempeqncnt}}
	%	\setcounter{equation}{5}
	% IEEE uses as a separator
	%\hrulefill
	% The spacer can be tweaked to stop underfull vboxes.
	%\vspace*{4pt}
    \vspace{-0.2cm}
\end{figure*}
\setcounter{mytempeqncnt}{\value{equation}}%
\begin{figure*}[t]
    \vspace{-0.5cm}
	% ensure that we have normalsize text
	\normalsize
	% Store the current equation number.
	%	\setcounter{mytempeqncnt}{\value{equation}}
	%\setcounter{equation}{4}
	% Set the equation number to one less than the one
	% desired for the first equation here.
	% The value here will have to changed if equations
	% are added or removed prior to the place these
	% equations are referenced in the main text.
	%	\setcounter{equation}{4}
		
	\begin{align}
    m\opt_n = \argmax_{m =1, \dots, \nums}\braces{c_{m,n}^t-\lambda \cdot \max_{\sfu\in\mathcal{U}} \, \tminrTht(\sfu, \vp_\sfu; \vs_m, \vg_n)  - \mu \cdot \max_{\sfu\in\mathcal{U}, \tau \in \mathcal{T}} \minr^\tau(\sfu, \vp_\sfu; \vs_m, \vg_n)  - \nu_m} \ \forall \ n
		% m\opt_n = \argmax_m \sum_{m=1}^{\nums}\braces{c_{m,n}^t-\lambda \cdot \max_{\sfu\in\mathcal{U}} \, \tminrTht(\sfu, \vp_\sfu; \vs_m, \vg_n)  - \mu \cdot \max_{\sfu\in\mathcal{U}, \tau \in \mathcal{T}} \minr^\tau(\sfu, \vp_\sfu; \vs_m, \vg_n)  - \nu_m} \ \forall \ n %\cdot x_{m,n}, 
    %,\forall n, 
    \label{eq:m_star}
	\end{align}
	% Restore the current equation number.
	%	\setcounter{equation}{\value{mytempeqncnt}}
	%	\setcounter{equation}{5}
	% IEEE uses as a separator
	\hrulefill
	% The spacer can be tweaked to stop underfull vboxes.
	%\vspace*{4pt}
    \vspace{-0.4cm}
\end{figure*}

\setcounter{equation}{26}
% \vspace{-0.4cm}

\subsection{Problem Formulation}
Having formulated a suitable time-averaged interference protection constraint, we now set our sights on optimizing the association made at handover time $t$, i.e., $\bX^t$.
In doing so, we assume the secondary system aims to maximize its own performance in terms of its capacity, though other measures could also be suitable.
In this vein, let us define the aggregate capacity across secondary users in cluster $\vg_n$ when served by secondary satellite $\vs_m$ throughout the handover period $\mathcal{T}$ as %$\mathcal{T}=[t,t+\Th)$ as
\begin{align}
    c_{m,n}^t 
    &\triangleq
    % \frac{1}{\Th} \frac{1}{|\mathcal{U}|}
    \sum_{\tau=t}^{t+\Th-1}\!\!
    \sum_{\sfv\in\mathcal{V}_{n,\ell}}
    \log_2\parens{1+\gamma^\tau(\sfv, \vs_m, \vc_{n,\ell})}
    \cdot y^\tau_{m,n,\ell}, % \\
    % &= 
    % \sum_{\tau=t}^{t+\Th-1}
    % \log_2\parens{\prod_{\sfv\in\mathcal{V}_{n,\ell}}\parens{1+\msinr^t(\sfv, \vs_m, \vc_{n,\ell}) \cdot y^t_{m,n,\ell}}}
\end{align}
where $\gamma^\tau(\sfv, \vs_m, \vc_{n,\ell})=\msinr^\tau(\sfv, \vs_m, \vc_{n,\ell})\,\ind{\vs_m\in\setsats^t_n}$,
% \begin{align}
%     \msinr^t_\mathrm{eff}(\sfv, \vs_m, \vc_{n,\ell})=\begin{cases}
%         &\msinr^t(\sfv, \vs_m, \vc_{n,\ell}), \ \text{if }\vs_m\in\setsats^t_n,\\
%         & 0, \ \text{otherwise}, 
%     \end{cases}
% \end{align}
% \begin{align}
%     \gamma^\tau(\sfv, \vs_m, \vc_{n,\ell})=
%         \msinr^\tau(\sfv, \vs_m, \vc_{n,\ell})\cdot\ind{\vs_m\in\setsats^t_n},
% \end{align}
which captures the minimum elevation requirement in \eqref{eq:min-elev}.
% of satellite $\vs_m$ relative cluster $\vg_n$. 
Treating this notion of sum capacity as our objective, our association problem at handover time~$t$ can be formulated by imposing the interference projection constraint derived before, along with aforementioned system constraints. % , as follows.
\begin{subequations}\label{eq:problem}
\begin{align}
\max_{\bX^t} \, & \sum_{m=1}^{\nums}\sum_{n=1}^{\numg}  c^t_{m,n} \cdot x^t_{m,n}
\label{eq:objective} \\
\st &  
\max_{\sfu\in\mathcal{U}} 
\sum_{m=1}^{\nums} \sum_{n=1}^{\numg}    	\tminrTht(\sfu, \vp_\sfu; \vs_m, \vg_n)  x_{m,n}^t \leq \ttminrth % \, \forall \, \sfu \in\mathcal{U}
\label{eq:avg_ipc_r} \\
& \max_{\substack{\sfu \in \mathcal{U}\\\tau \in \mathcal{T}}}\sum_{m=1}^{\nums} \sum_{n=1}^{\numg} \minr^\tau(\sfu, \vp^t_\sfu; \vs_m, \vg_n)x_{m,n}^t \! \leq \! \minrth^{\mathrm{max}}%\forall \ \tau \in [t,t+\Th), \sfu \in \mathcal{U} 
\label{eq:abs_ipc_r}\\
% & \sum_{m=1}^{\nums} \sum_{n=1}^{\numg} \minr^\tau(\sfu, \vp^t_\sfu; \vs_m, \vg_n)x_{m,n}^t  \leq \minrth^{\mathrm{max}} \ \forall \sfu, % \in \mathcal{U}, 
% \tau \in \mathcal{T} %\forall \ \tau \in [t,t+\Th), \sfu \in \mathcal{U} 
% \label{eq:abs_ipc_r}\\
& \sum_{n=1}^{\numg} x^t_{m,n} \leq 1 \ \forall \ m \label{eq:max_num_clusters} \\
&\sum_{m=1}^{\nums} x^t_{m,n} \leq 1 \ \forall \ n % , \ \ 
\label{eq:single_association} \\
% \, 
& x^t_{m,n} \in \braces{0,1} \ \forall \ m, n\label{eq:integer}
\end{align}
\end{subequations}
%where $\vs_m\in\mathcal{S}^t_{n}$
%\blue{Shoud we separately define $\tminrth$ and/or $\tminrTht$, as we don't have $\tminrth$ explicitly in the problem. What do you think?}\\
Solving problem \eqref{eq:problem} at handover time $t$ would yield the secondary system satellite-cluster association $\bX^t$ that maximizes the sum capacity throughout the handover period $[t,t+\Th)$ while ensuring the average and absolute interference inflicted upon any given primary user are constrained to some specified thresholds.
Notice that maximizing the secondary system sum capacity amounts to choosing the association $\bX^t$ that delivers high {\sinr} to its users and yields lower received interference inflicted by the primary system.
In other words, the optimal secondary satellite serving a given cluster of secondary users would likely be spatially separated from the primary satellites serving nearby primary users, in terms of both its own capacity and interference protection. 
It is important to note that, if the interference constraints are particularly strict, the secondary system may be unable to serve certain clusters during a given handover period.
Such outage events are an undesired yet natural consequence of overly protective interference constraints.

%\vspace{-0.25cm}
\subsection{Lagrangian Relaxation and Subgradient Iterations} 
\setcounter{equation}{29}
Problem \eqref{eq:problem} is a matching problem \cite{lmckp_david}, known to be NP-hard \cite{np_hard}. 
In light of this, we solve it by relaxing inequality constraints \eqref{eq:avg_ipc_r}, \eqref{eq:abs_ipc_r}, and \eqref{eq:max_num_clusters} through Lagrangian relaxation \cite{lagrangian_relax}. 
% \ipr{LET'S DISCUSS THIS.}
% Lagrangian relaxation is a technique for solving constrained optimization problems by simplifying the problem through the relaxation of inequality constraints, which are incorporated into the objective function using Lagrangian multipliers \cite{lagrangian_relax}. 
This transforms the original problem into a dual problem which can be decomposed into smaller, independent subproblems that are easier to solve using dynamic programming. 
The Lagrangian multipliers can then be optimized via subgradient iteration \cite{integer_prog}, i.e., iteratively updated based on the violation of the original constraints. % , and the dual function, representing the relaxed problem, is optimized to provide bounds on the original problem's solution \cite{lagrangian_relax}. 

Let ${\lambda}\in \mathbb{R} $, ${\mu}\in \mathbb{R}$, and $\boldsymbol{\nu}\in\mathbb{R}^{\nums \times 1}$ be Lagrange multipliers and define $g(\bX^t, {\lambda},\mu, \boldsymbol{\nu})$ as in \eqref{eq:ld}.
% \begin{multline}
% g(\bX^t, \lambda, \mu, \boldsymbol\nu)
% =
% \sum_{m=1}^{\nums} \sum_{n=1}^{\numg}
% \parens{c^t_{m,n} 
% - \lambda \cdot \max_{\sfu\in\mathcal{U}} \, \tminrTht(\sfu, \vp_\sfu; \vs_m, \vg_n)  
% -\mu \cdot \max_{\sfu\in\mathcal{U}, \tau \in \mathcal{T}} \minr^\tau(\sfu, \vp_\sfu; \vs_m, \vg_n)
% - \nu_m
% } \cdot x^t_{m,n} \\
% + \lambda \cdot \tminrtheff^{t}  
% +\mu\cdot \tminrth^{\mathrm{max}} 
% + \sum_{m=1}^{\nums}\nu_m +  \label{eq:ld}
% \end{multline}
% \red{It should be $\min\parens{\ttminrth, \minrth}$ instead of \tminrth. Can we have a good notation for $\min\parens{\ttminrth, \minrth}$? I'm trying to come up with, but...  }
Due to the inequality constraints \eqref{eq:avg_ipc_r}, \eqref{eq:abs_ipc_r}, and \eqref{eq:max_num_clusters}, 
% the dual function $g(\lambda,  \mu, \boldsymbol\nu) = \max_{\bX^t} g(\bX^t; \lambda, \mu, \boldsymbol\nu)$ is the upper bound on the optimal value of the original integer problem for all for all $\lambda\geq0$, $\mu\geq0$, and $\boldsymbol\nu\geq\boldsymbol0$.
we have 
\begin{align}
\max_{\bX^t} \,  g(\bX^t, {\lambda}, \mu, \boldsymbol{\nu}) 
\geq 
\max_{\bX^t} \,\sum_{m=1}^{\nums}\sum_{n=1}^{\numg} c^t_{m,n} \, x^t_{m,n},  
\label{eq:lr_upper}
\end{align}
for all $\lambda\geq0$, $\mu\geq0$, and $\boldsymbol\nu\geq\boldsymbol0$.
%Expression \eqref{eq:lr_upper} indicates that our Lagrangian relaxation is an upper bound on the optimal value of the original integer problem. Therefore: 
% \begin{subequations}
% \begin{align}
% \min_{\lambda \geq 0,  \mu\geq 0,  \boldsymbol\nu \geq \boldsymbol0} 
% \, \max_{\bX^t} \ & g(\bX^t, \lambda, \mu, \boldsymbol\nu) \\
% \st 
% & x^t_{m,n} \in \braces{0,1} \ \forall \ m, n \label{eq:zerone} \\
% & \sum_{m=1}^{N_s} x^t_{m,n} \leq 1 \ \forall \ n. \label{eq:zerone2}
% \end{align}
% \end{subequations}
The dual function $g(\lambda, \mu, \boldsymbol\nu)$ is obtained by maximizing the Lagrangian over $\bX^t$: %subject to the assignment constraints: 
\begin{subequations}
\begin{align}
g(\lambda, \mu, \boldsymbol\nu) 
\triangleq \max_{\bX^t} \ & g(\bX^t; \lambda, \mu, \boldsymbol\nu) \label{eq:dual_p} \\
\st 
& \sum_m x_{m,n}^t \leq 1, 
\,\ x_{m,n}^t \in \braces{0,1}. \label{eq:zerone}
\end{align}
\end{subequations}
Note that Lagrangian relaxation with $\braces{0,1}$ constraints in  \eqref{eq:zerone} preserves the optimality of the original problem \cite{lagrangian_relax}.
%\blue{The optimal solution of the relaxed problem is proven to be optimal due to the $\braces{0,1}$ constraints in \eqref{eq:zerone} and \eqref{eq:zerone2} \cite{lagrangian_relax}. %Decomposing the dual problem into cluster-level subproblems allows selecting the satellite that maximizes the objective for given Lagrange multipliers. % 
%}
% The dual function $g(\lambda, \mu, \boldsymbol\nu)$ is obtained by maximizing the Lagrangian over $\bX^t$ subject to the assignment constraints: 
% \begin{subequations}
% \begin{align}
% 	g(\lambda, & \mu, \boldsymbol\nu) = \max_{\bX^t} g(\bX^t; \lambda, \mu, \boldsymbol\nu) \\
% 	\st & \sum_m x_{m,n}^t \leq 1, \\
% 	&\,\, x_{m,n}^t \in \braces{0,1}.
% \end{align}
% \end{subequations}
\setcounter{equation}{34}
%The Lagrange multipliers are then optimized via subgradient iterations.

Given the Lagrangian multipliers, % $\lambda$, $\mu$ and $\boldsymbol\nu$, 
%the maximization over $x_{m,n}^t$ 
the dual function \eqref{eq:dual_p} can be decomposed into %computationally efficient 
independent cluster-level subproblems as in \eqref{eq:sp_1}. Thus, at each cluster $n$, we find the optimal satellite $m_n\opt$ satisfying \eqref{eq:m_star}: %-$\eqref{eq:sp_3}. 
%We aim to find the optimal $m_n\opt$ satisfying \eqref{eq:m_star}, 
if $m = m\opt_n$ we have $x_{m,n}^t=1$, otherwise  $x_{m,n}^t=0$, for each $n$. 
% \begin{align}
% 	x_{m,n}^t = \begin{cases}
% 		1, \qquad \text{if } m = m\opt_n, \\
% 		0, \qquad \text{otherwise}.
% 	\end{cases}
% \end{align}
%Note that each subproblem always yields an optimal integer solution $x_{m,n}^t$ for each cluster $\vg_n$. 
%By decomposing the original problem \eqref{eq:problem} into cluster-level subproblems \eqref{eq:sp_1} through the relaxation of three inequality constraints, %\eqref{eq:avg_ipc_r}, \eqref{eq:abs_ipc_r}, and \eqref{eq:max_num_clusters}, 
%we transform the problem into a computationally efficient form.   
%enabling it to be effectively used for large-scale problems.
%We then compute the dual function $g(\lambda, \mu, \boldsymbol\nu)$ by plugging in the selected $x_{m,n}^t$, and we use subgadient methods to solve the dual problem $\min_{\lambda \geq 0, \mu\geq 0, \boldsymbol\nu \geq \boldsymbol{0}} g(\lambda, \mu, \boldsymbol\nu)$. 
% \begin{align}
%     \min_{\lambda \geq 0, \mu, \boldsymbol\nu \geq 0} g(\lambda, \mu, \boldsymbol\nu).
% \end{align}
%In Lagrangian relaxation methods, 
%Subgradient iterative techniques \cite{integer_prog} can be used to update the set of Lagrange multipliers. 
The Lagrangian multipliers can then be optimized by subgradient iteration \cite{integer_prog}, %i.e., they are iteratively updated based on the violation of the original constraints, 
and the dual problem $\min_{\lambda \geq 0, \mu\geq 0, \boldsymbol\nu \geq \boldsymbol{0}} \, g(\lambda, \mu, \boldsymbol\nu)$ is optimized to provide bounds on the original problem's solution \cite{lagrangian_relax}. 
% During the subgradient iteration, certain associations $x_{m,n}^t$ are set to zero if they would otherwise violate the original constraints. 
% As a result, some clusters may end up in outage.
This concludes our proposed secondary satellite selection mechanism, which we thoroughly assess in the next section.

\comment
{
	\subsection{Linear Relaxation}
	The optimization problem in the previous section is a form of NP-hard multi-choice knapsack problem. The MCKP is a variant of the classic knapsack problem. MCKP involves a set of items, each possessing a weight and a value. The items are divided into several disjoint classes. The objective of MCKP is to select exactly one item from each class to maximize the total value while adhering to a predefined weight limit.
	In this work, the secondary ground cells and satellites are modeled as classes and items, respectively, such that each secondary cell should select one secondary satellite within the set $\setsats$. 
	Over the past several decades, numerous methods for solving MCKP have been proposed, and we will solve the problem using dynamic programming, and Lagrangian. 
	
	For the MCKP, a common approach to reduce the cost of solving the NP-hard problem is \textit{linear programming relaxation} in which the integral constrain \eqref{eq:single_associations} is removed by replacing \eqref{eq:integer} with: 
	\begin{align}
		0 \leq x_{i,j} \leq 1, \forall i, j.
	\end{align}
	This relaxed problem is often called the \textit{linear} MCKP or LMCKP \cite{lmckp_david} and it is relatively easy to solve. An interesting property of LMCKP is that the solution that maximizes the objective function \eqref{eq:objective} represents an upper bound on the MCKP from which it derives. This upper bound is also at least as tight as an upper bound resulting from Lagrangian relaxation and is potentially better \cite{bit_alloc_mckp}.
	
	%It has been proved that \eqref{eq:objective} is optimal when $x_{i,j}$ is either $1$ or $0$ \cite{}, so we can relax the first integer constraint to 
	%\begin{align}
	%	0 \leq x_{i,j} \leq 1, \forall i, j 
	%\end{align}
}

%\section{Numerical Results}
\section{Evaluating Our Secondary\\Satellite Selection Mechanism}\label{sec:numerical_results}

To conduct a thorough and relevant analysis on the in-band coexistence of two dense \leo satellite communication systems under our proposed selection mechanism, we consider two preeminent commercial systems at $20$~GHz: %Starlink by SpaceX and Project Kuiper by Amazon. We consider 
Starlink as the primary and Kuiper as the secondary system; this is motivated by the fact that Starlink has priority rights over Kuiper to transmit downlink in the $19.7$--$20.2$~GHz band. %  for downlink transmission.
%Indeed, based on current regulations, Kuiper is permitted to operate within that band along with Starlink with the understanding that it will not cause ``prohibitive interference" to Starlink---coinciding with the analyses herein. 
We simulate the Starlink and Kuiper constellations in a Walker-Delta fashion \cite{wd}, {using \textsc{Matlab}'s Aerospace Toolbox \cite{matlab}}, based on the orbital parameters detailed in \tabref{tab:spaceX} and \tabref{tab:kuiper}, which are extracted from public filings \cite{kuiper, spaceX_req, spaceXX}. 
The total numbers of primary and secondary satellites are correspondingly $\nump=6900$ %\footnote{We consider Starlink's Gen1 satellites and part of Gen2 satellites.}
and $\nums=3236$. 
Primary and secondary satellite locations are sampled every $0.1$ sec in simulation. 
%We sample primary and secondary satellite locations  every $0.1$ seconds, select serving satellites for the considered ground area, and calculate the \ginr and \gsnr of ground users for each system. 
Although $\Th$ and $\Tw$ were defined in the previous section in units of time samples, they are expressed here in seconds to provide a more meaningful  interpretation.

\begin{table}[t]
	% \vspace{-0.1cm}
	\caption{SpaceX's Starlink Constellation Parameters \cite{spaceX_req,spaceXX}}
	\centering
	\label{tab:spaceX}
	\begin{tabular}{| c| c|c|c|c| }
		\hline
		% Alt. ($km$) & Inc. (${}^\circ$)  & $\mathrm{N_{pl}}$ & $\mathrm{N_{sat-pl}}$ & $\mathrm{N_{sat-total}}$  \\
		Altitude & Inclination &  Planes & Satellites/Plane & Total No.~Satellites \\
		\hline
		$540$ km & $ 53.2^\circ$ &$ 72$ & $22$& $1584$ \\
		\hline
		$550$ km & $ 53^\circ$ & $72$ & $22$ & $1584$ \\
		\hline 
		$560$ km & $97.6^\circ$ & $4$ & $43$ & $172$ \\
		\hline 
		$560$ km & $97.6^\circ$ & $6$ & $58$ & $348$ \\
		\hline
		$570$ km & $70^\circ$ &$ 36$ &$ 20$ &$ 720$ \\
		\hline
		$530$ km & $33^\circ$ & $28$ & $89$ & $2492$ \\
		\hline
	\end{tabular}
	\vspace{-0.3cm}
\end{table}

\begin{table}[t]
	\caption{Amazon's Project Kuiper Constellation Parameters \cite{kuiper}}
	\centering
	\label{tab:kuiper}
	\begin{tabular}{| c| c|c|c|c| }
		\hline
		Altitude & Inclination &  Planes & Satellites/Plane & Total No.~Satellites \\
		\hline
		$590$ km & $33^\circ$ & $28$ & $28$ &$ 784$ \\
		\hline 
		$610$ km & $42^\circ$ & $36$ & $36$ & $1296$ \\
		\hline 
		$630$ km & $51.9^\circ$ & $34$ & $34$& $1156$ \\
		\hline
	\end{tabular}
	\vspace{-0.3cm}
\end{table}
% multiple antenna panels, utilizing 
Each satellite is equipped with 64$\times$64 phased array antennas capable of generating multiple spot beams to serve ground users, delivering a maximum transmit beam gain of $36$~dBi, and the receive antenna of ground users are modeled with 32$\times$32 phased array antennas. 
% Based on the $-3$ dB transmit beam contours. 
Based on the resulting $-3$~dB transmit beam contours, the radius of ground cells is approximately $10$~km, and we consider $\numg=10$ clusters as illustrated in \figref{fig:geo_map}, each comprised with $\numc=127$ ground cells. 
These values are on par with their public filings \cite{spaceX_ss, kuiper_ss}. 
% The number in each hexagon in \figref{fig:geo_map} represents a form of prioritization across clusters that dictates the order in which clusters are assigned their serving satellites. %  in the order of satellite assignments.
% In other words, clusters may have different priorities based on factors such as the number of subscribers. 
% For example, given the limited number of satellites in orbit, 
% For example, this may capture the fact that denser areas may be prioritized for higher capacity.
% This prioritization is purely for the sake of defining an explicit satellite selection policy (as we will describe shortly) and has limited impact on the results that follow.
% \ipr{Mention that this is only relevant in the next section?}
We evaluate coexistence of the two satellite systems for a various number of spot beams per satellite, exploring four configurations: 8, 16, 24, and 32 beams \cite{leo_comparison,num_spotbeams}. 
We consider a frequency reuse factor of three, following \cite{kuiper}. % \red{per frequency band.}
% As we are interested in the worst-case interference, 
To assess a worst-case interference scenario, we assume that both the primary and secondary systems share the same ground cell and cluster deployment  with overlapping cell centers, % , as discussed in \secref{sec:methodology}, 
and users are uniformly distributed across all ground cells. % , each positioned at the center of its respective cell. 
%To assess worst-case interference, 
To comply with power flux density regulations \cite{cfr25} set by regulatory authorities such as the FCC, a simple transmit power control mechanism is implemented at each satellite based on path distance and elevation angle. 
Specifically, the maximum \eirp provided in \tabref{tab:sim_parameters} applies only to satellites operating at nadir at the highest altitude in the constellation. 
%For satellites at lower altitudes, the \eirp is adjusted to account for reduced path loss, ensuring compliance with power flux density limits \ipr{imposed by higher-altitude satellites IS THIS PHRASE NEEDED?} \cite{cfr25}.
Despite the high velocity of LEO satellites, which induces significant Doppler effects, we do not explicitly account for such in our analysis, as they are accurately estimated and compensated for in practice based on satellite orbits and user locations \cite{doppler1,doppler2}.
% Doppler effects in real systems can be accurately estimated and compensated for \cite{doppler1,doppler2} using readily available information on satellite orbits and ground user locations.
Key simulation parameters are summarized in \tabref{tab:sim_parameters}, all of which are based on public filings and 3GPP specifications~\cite{spaceX_ss,kuiper_ss,3gpp38821}.

\begin{table}[t]
	\vspace{-0.2cm}
	\caption{Key Simulation Parameters \cite{spaceX_ss,kuiper_ss, 3gpp38821}}
	\vspace{-0.12cm}
	\centering
	\label{tab:sim_params}
	\begin{tabular}{|c|c|c|}
		\hline
		\multicolumn{3}{|c|}{Primary and Secondary Satellites}\\
		\hline
		\multirow{3}{*}{Tx. Antenna Array} & No. Antennas &{64$\times$64}\\
		\cline{2-3}
		&3~dB Beamwidth& {$1.6^\circ$} \\
		\cline{2-3}
		& Max. Beam Gain &{$36$~dBi}\\
		\hline
		\multirow{2}{*}{Max. EIRP} & Primary Sat. & {$-54.3$~dBW/Hz}\\
		\cline{2-3}
		& Secondary Sat. & {$-53.3$~dBW/Hz}\\
		\hline
		%\multicolumn{2}{|c|}{No. Spot Beams, $\numbeams$} & $8$, $16$, $24$, or $32$\\
		%\hline
		\multicolumn{3}{|c|}{Primary and Secondary Ground Users}\\
		\hline
		\multirow{3}{*}{Rx. Antenna Array} & No. Antennas   & 32$\times$32 \\
		\cline{2-3}
		& Max. Beam Gain  &$30$~dBi \\
		\cline{2-3}
		& 3~dB Beamwidth &$3.2^\circ$\\
		\hline
		\multicolumn{2}{|c|}{Noise Power Spectral Density} & {$-174$~dBm/Hz}\\ 
		\hline
		\multicolumn{2}{|c|}{Noise Figure} & {$1.2$~dB}\\ 
		\hline
		
	\end{tabular}
	\vspace{-0.4cm}
	\label{tab:sim_parameters}
\end{table}

%\vspace{-0.2cm}
\subsection{Satellite-to-Cluster Association of Primary System}
%The satellite-to-cluster association depends on the handover policy adopted by the satellite systems. 
A natural starting point in our evaluation is to assess system performance when the secondary system makes no attempt to protect the primary system.
It is thus necessary to define the association policies initially employed by both systems.
(Afterwards, the secondary system association will be governed by our proposed satellite selection mechanism.)
In this vein, we consider two plausible handover policies: highest elevation (HE) and maximum contact time (MCT) \cite{sat_handover}. 
Under the HE policy, a given satellite system selects the satellite with the highest elevation angle relative to the cluster it aims to serve. {In this case, the handover period is set to $\Th = 15$ sec, coinciding with that observed in Starlink~\cite{starlink_measure}. }
In contrast, the MCT policy prioritizes the satellite that offers the longest visible time above the minimum elevation angle. 
{In this approach, a handover is initiated when the serving satellite is expected to descend below the minimum elevation angle, transitioning to the satellite that provides the longest remaining visibility.} 
In applying these policies cluster-by-cluster, the prioritization numbering shown in \figref{fig:geo_map} is used to dictate the order in which this association is carried out.\footnote{This prioritization is simply used to define an explicit satellite selection policy and does not have significant influence on the results that follow.}
This can capture prioritizing denser areas for higher capacity. 
Under the HE policy, if a particular satellite is that with the highest elevation across multiple clusters, it is assigned to the highest priority cluster. 
This prioritization is analogously applied to the MCT policy.
\subsection{Independent Operation of the Two Systems}
% A \ipr{usual WORD CHOICE} starting point is to 
We begin our evaluation by assessing the performance 
% of two independent primary and secondary satellite systems 
when the secondary system operates without any attempt to protect the primary system. 
To do so, at each time step in our simulation (under a 0.1 sec resolution), we compute the \ginr and \gsinr of both primary and secondary users after performing the aforementioned satellite-to-cluster association policies.
Note that the only interference considered, for the sake of clarity, is that inflicted by the other system.
In reality, intra-system interference would also be present but may be carefully controlled by each constellation separately, and we therefore omit it to focus exclusively on inter-system interference.
% determine the satellite-to-cluster association of each system depending on the particular handover policy, and then compute the \ginr and \gsinr of both primary and secondary users. %, assuming both systems have complete knowledge about each other's satellite-to-cluster.  
% At every time step (under a 0.1 second resolution), we determine the satellite-to-cluster association of each system depending on the particular handover policy, and then compute the \ginr and \gsinr of both primary and secondary users. %, assuming both systems have complete knowledge about each other's satellite-to-cluster.  

%For the evaluation of this work, the frame time of the secondary system is offset by $n_0=2$ samples, corresponding to $0.2$ seconds, from that of the primary system.

% \ipr{THE PRIMARY INR SHOWN IS ONLY THAT CONTRIBUTED BY THE SECONDARY RIGHT? Yes, says eskim.}

\begin{figure}[t]
    \vspace{-0.2cm}
    \centering    \includegraphics[width=\linewidth,height=\textheight,keepaspectratio]{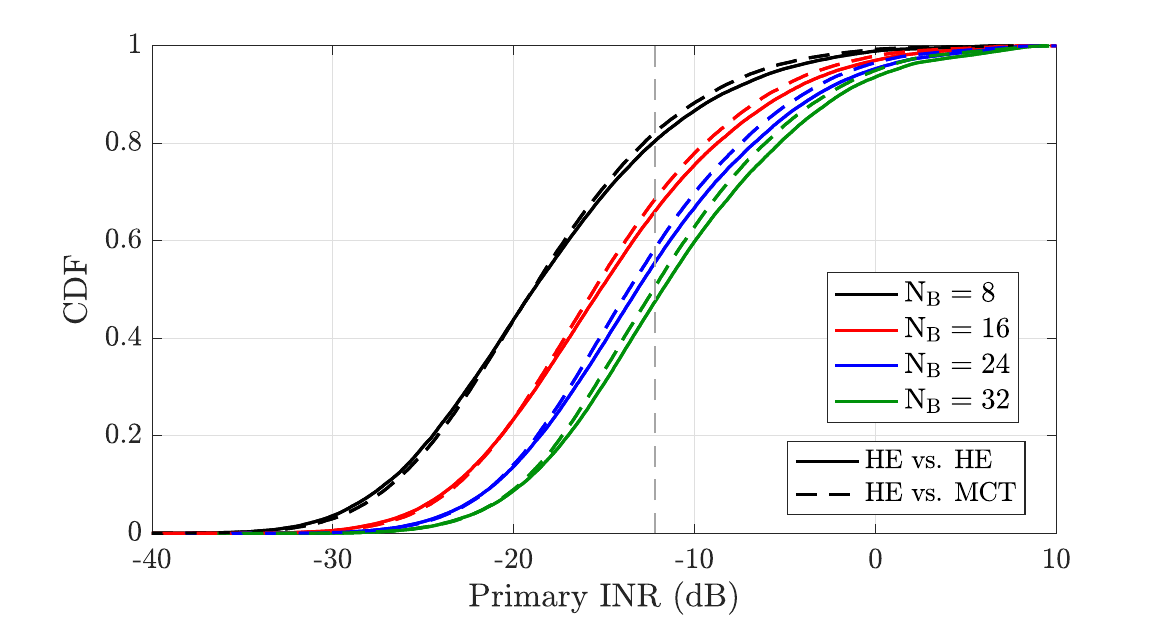}
    \caption{The \cdf of \ginr across primary users and across time for a various number of spot beams and handover policies. The \ginr at approximately $20\%$ of primary users exceeds $-12.2$~dB when $\numbeams=8$, further increasing with $\numbeams$. The upper tail motivates the need for the secondary system to explicitly protect primary users.}
    \label{fig:cdf_multi_system_pinr}
   \vspace{-0.4cm}
\end{figure}

We examine the primary user \ginr for a various number of spot beams per satellite and assess performance under the aforementioned handover policies of HE and MCT.
% whether different handover policies can aid coexistence.
\figref{fig:cdf_multi_system_pinr} depicts the \cdf of the \ginr across primary users and across time, representing the aggregate interference from all active satellites and spot beams of the secondary system, i.e., $\minr^t(\sfu, \vp)$ as defined in \eqref{eq:pinr_def}. 
% The primary system is modeled to perform handovers every $15$ seconds across all clusters simultaneously. % , and the secondary system adopts the same cluster p%, starting from the first time instance in the simulation.
% Both primary and secondary satellites are assumed to have the same number of spot beams per satellite. 
Solid lines in the figure represent scenarios where the secondary system adopts the same handover policy as the primary system, and dotted lines indicate cases where the secondary system differs in its handover policy, using MCT. 
We can see that adopting a different handover policy at the secondary system slightly reduces the \ginr inflicted onto primary users. % , but the improvement is minimal.
This marginal improvement is presumably due to improved spatial separation of the selected primary and secondary satellites, as MCT naturally tends to select satellites earlier in their overhead pass and thus closer to the horizon, whereas HE selects those closer to nadir.

\begin{takeaway}[Independent operation of the two systems leads to prohibitively high interference]
As evident in \figref{fig:cdf_multi_system_pinr}, even with only $\numbeams=8$ spot beams per satellite, there is substantial interference inflicted onto primary users a non-negligible fraction of the time.
Referencing an \ginr threshold of $-12.2$~dB, as widely employed by the ITU \cite{ntia_ipc,itur1323}, we can see that around $20\%$ of primary users would exceed this interference threshold.
This naturally worsens as the number of spot beams increases.
With $\numbeams=32$, this \ginr threshold is exceeded for around $50\%$ of users, regardless of the policies employed by the two independent systems.
These results motivate the need for mechanisms to explicitly protect primary users, in order to facilitate a healthy coexistence paradigm between two dense \leo satellite constellations.
 %\vspace{-0.12cm}
\end{takeaway}

\vspace{-0.2cm}
\subsection{Coexistence Under Our Proposed Approach}

\begin{figure}[t!]
\vspace{-0.2cm}
	\centering
\includegraphics[width=\linewidth,height=\textheight,keepaspectratio]{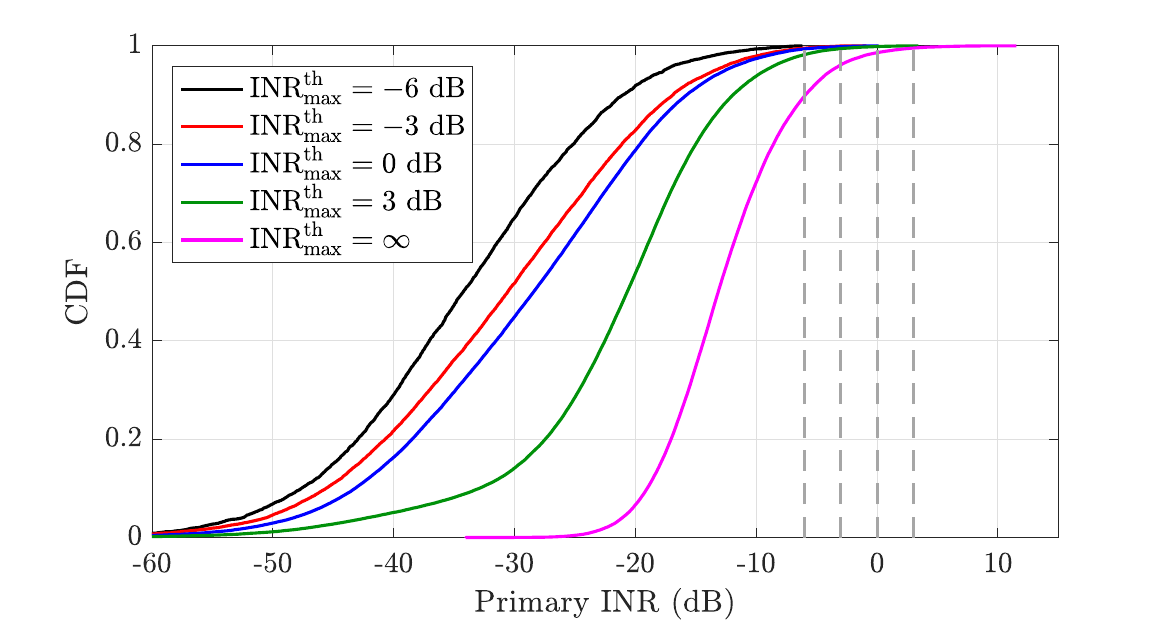}
	\caption{The \cdf of the primary user \ginr (across users and time) for various \minrthmax values under our proposed approach. The average \ginr threshold \tminrth is set to $-6$ dB, with $\Th = 15$ sec and $\Tw = 10$ sec.}
		\label{fig:lr_pinr_ssinr_a}
        \vspace{-0.4cm}
\end{figure}

\begin{figure}[t!]
    %\vspace{-0.3cm}
	\centering
\includegraphics[width=\linewidth,height=\textheight,keepaspectratio]{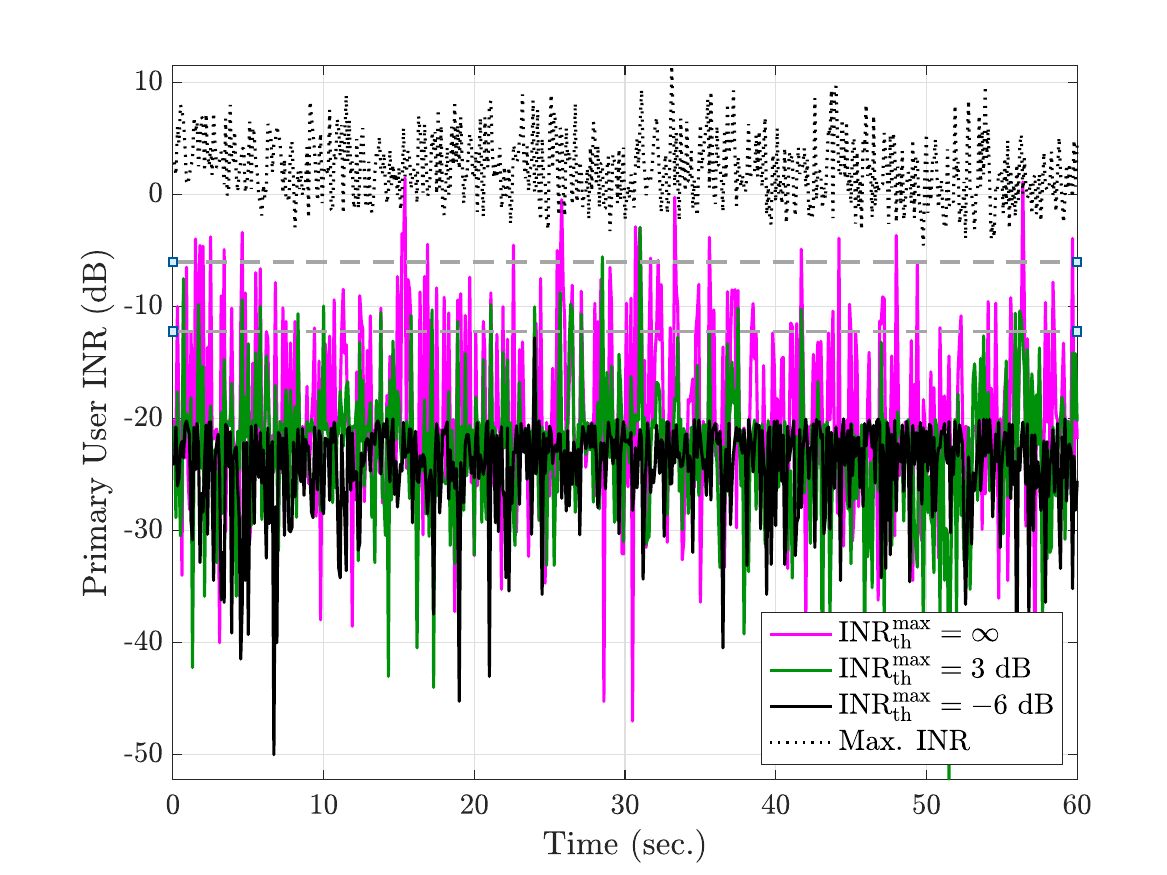}
	\caption{Instantaneous \ginr experienced by a single primary ground user during a $60$-sec period under our proposed scheme for various absolute thresholds $\minrth^\mathrm{max}$. The dotted line indicates the maximum \ginr across all primary ground users throughout the handover period. The time-average \ginr constraint is set to $\tminrth = -6$~dB, with $\Th = 15$ sec and $\Tw = 10$ sec.}
		\label{fig:inst_inr_max}
        \vspace{-0.4cm}
\end{figure}

Having motivated the need for explicit protection mechanisms, we now turn our attention toward assessing our proposed approach described in \secref{sec:sat_selection}, with $\numbeams=16$, the typical number of spot beams per satellite in Starlink~\cite{num_spotbeams}.
In \figref{fig:lr_pinr_ssinr_a}, we plot the \gcdf of the primary \ginr (across users and time) when employing our protection mechanism with absolute protection thresholds $\minrthmax$, where the time-average threshold has been set to $\tminrth = -6$~dB.
The black line with $\minrthmax = -6$~dB is thus equivalent to the strict protection constraint described in \eqref{eq:strict_ipr}.
In this case, we can see that indeed the primary \ginr never exceeds $-6$~dB and is orders of magnitude below $-6$~dB most of the time.
Upon increasing the absolute threshold $\minrthmax$, the distribution shifts rightward but still lies well below $-6$~dB most of the time. 
The upper tail exceeds $-6$~dB, however, as a consequence of the flexibility provided by increasing $\minrthmax$.
Since the time-average threshold remains at $\tminrth = -6$~dB, the distribution is nonetheless constrained to remain mostly below $-6$~dB.
We can see that, even with $\minrthmax = 3$~dB, the primary \ginr only exceeds $-6$~dB about 3\% of the time.
If an instantaneous \ginr of $-12.2$~dB were used to define protection, we can see that a time-average threshold of $\tminrth = -6$~dB does meet this a high percentage of the time for a suitably chosen $\minrthmax$.
% The upper tail has inched rightward, however, 
This illustrates that, by pairing an absolute interference constraint with a time-averaged one, our proposed protection mechanism is capable of meeting more flexible yet still protective operating points for appropriately chosen $\tminrth$ and $\minrthmax$.
Soon, we will show that this flexibility greatly improves secondary system performance. 
% What remains to fully justify this, however, is to gauge secondary system performance, which we do shortly. 

Thus far, we have evaluated the performance of our proposed approach in terms of the \gcdf of \ginr across all primary users and time. 
To gain clearer insight into our approach's behavior, we now examine the instantaneous \ginr of a single primary user over 60 sec (four handover periods).
This is shown in \figref{fig:inst_inr_max} for various absolute thresholds $\minrthmax$ with the time-averaged \ginr constraint $\tminrth=-6$~dB, when $\Tw=10$ sec and $\Th=15$ sec. 
Notice that with $\minrthmax = -6$~dB, the \ginr at this particular user remains well below the time-average constraint of $-6$~dB throughout the entire handover period; there is only one narrow spike that approaches $-12.2$~dB.
Increasing to $\minrthmax = 3$~dB leads to occasional, short-lived spikes in \ginr exceeding $-12.2$~dB.
% Note that, when $\minrthmax = 3$~dB, 
However, this particular user only sees a worst-case \ginr around $-5$~dB and \ginr only exceeds the time-average threshold of $\tminrth = -6$~dB twice in this minute-long duration. 
% a higher spikes in \ginr, with relatively infrequent  instances exceeding $-12.2$~dB.
When $\minrthmax = \infty$, \ginr far more frequently exceeds both $-12.2$~dB and $-6$~dB and occasionally approaches or exceeds $0$~dB.
% is unbounded but not drastically, as the maximum \ginr remains around $2$~dB for this particular user.
For context, we have included the maximum \ginr across users when $\minrthmax=\infty$ in the dotted line, which illustrates that there are indeed users across the network that do incur prohibitively high \ginr at virtually any given time.
However, the interference profile of this single user suggests that high \ginr at any one user is typically short-lived. 
% Furthermore, we can see that the worst-case \ginr across users is at most around $10$~dB, which is constrained by the system's need to satisfy the time-averaged protection constraint of $-12.2$~dB.
This motivates us to quantify the fraction of primary users that see high \ginr at any given time next, but let us first draw the following conclusions.

% over time that primary users see high.

% \begin{figure}[t!]
%     \vspace{-0.3cm}
% 	\centering
% \includegraphics[width=\linewidth,height=\textheight,keepaspectratio]{nn_fig/inst_primary_inr_r6}
% 	\caption{Instantaneous \ginr experienced by a single primary ground user during a $60$-sec period under our proposed scheme for various absolute thresholds $\minrth^\mathrm{max}$. The dotted line indicates the maximum \ginr across all primary ground users throughout the handover period. The time-average \ginr constraint is set to $\tminrth = -6$~dB, with $\Th = 15$ sec and $\Tw = 10$ sec.}
% 		\label{fig:inst_inr_max}
%         \vspace{-0.4cm}
% \end{figure}

%\vspace{-0.12cm}
\begin{takeaway}[For a fixed satellite association, primary user \ginr can fluctuate rapidly and widely throughout a handover period]
This can be attributed to the many satellites and beams that comprise this complex, time-varying interference setting, further illustrating the challenges in meeting a strict protection constraint throughout an entire handover period.
Furthermore, increasing (or even removing) the absolute \ginr constraint does not necessarily result in \textit{all} users seeing extremely high \ginr. 
As we will see, this relaxation can greatly improve utilization of the secondary system by allowing it to select satellites which would otherwise be unable to serve particularly problematic primary users. % , which are a product of this extremely complex interference scenario.   
%\vspace{-0.12cm}
\end{takeaway}

While raw primary user \ginr is certainly important, it is only one component in assessing coexistence.
For this reason, we introduce two metrics to more straightforwardly assess coexistence under our proposed satellite selection mechanism:
\begin{itemize}
    \item[i.] the violation rate, defined as the fraction of primary users whose instantaneous \ginr exceeds the time-average threshold \tminrth, i.e., $\frac{1}{|\mathcal{U}|} \sum_{\sfu \in \mathcal{U}} \ind{\minr^t(\sfu,\vp) > \tminrth}$;
    \item[ii.] secondary system utilization, defined as the fraction of secondary ground users actively served by the secondary system, i.e., $\frac{1}{\numg} \sum_{m=1}^\nums \sum_{n=1}^\numg x^t_{m,n}$.
\end{itemize}
We seek a low violation rate and a high utilization for healthy coexistence, both of which depend on the time-average constraint $\tminrth$ and  the absolute constraint $\minrthmax$.
Note that the violation rate will be zero when $\minrthmax = \tminrth$, but this may lead to poor utilization for strict $\minrthmax$, as the secondary system may be unable to serve its users while guaranteeing protection, forcing it to leave some clusters un-served in a given handover period.

\begin{figure}[t!]
    \vspace{-0.2cm}
	\centering
\includegraphics[width=\linewidth,height=\textheight,keepaspectratio]{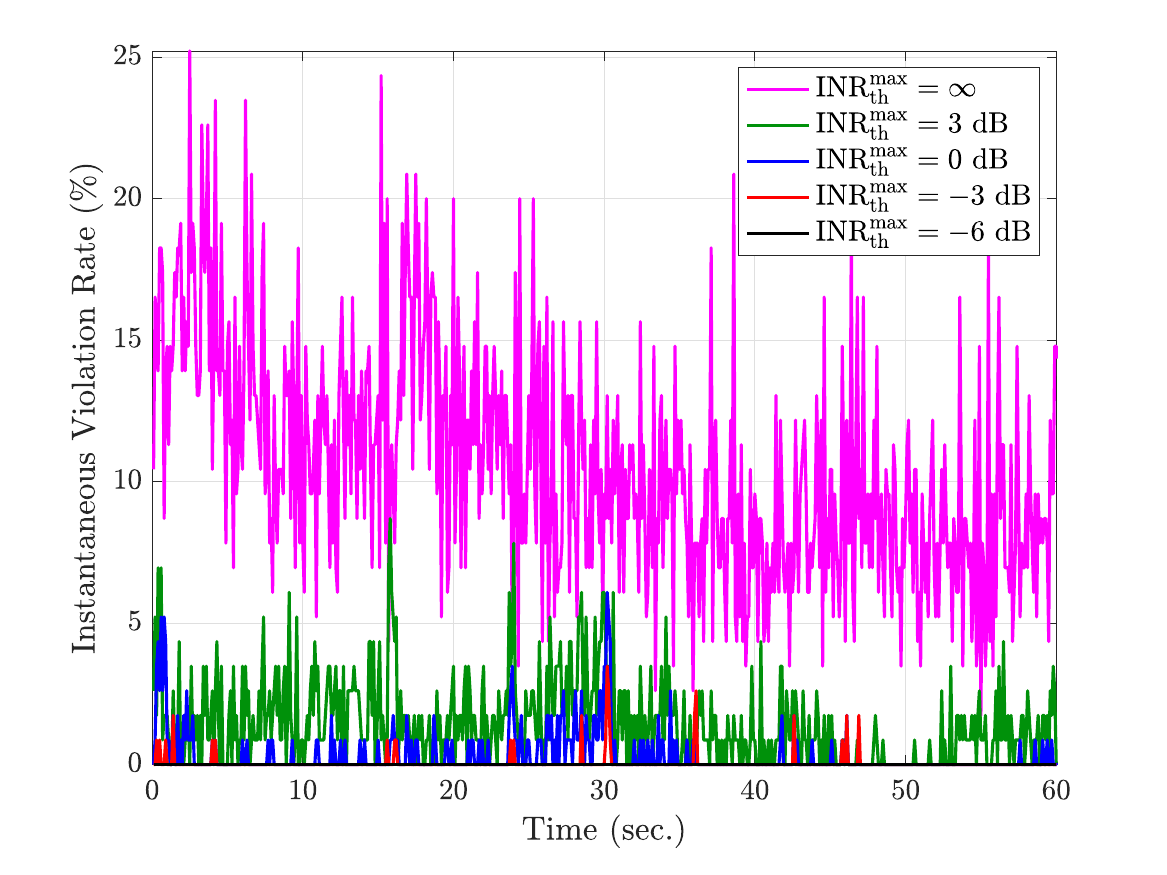}
	\caption{Instantaneous violation rate (across primary users) for various absolute thresholds $\minrthmax$, where $\tminrth = -6$~dB, $\Th = 15$ sec, and $\Tw = 10$ sec.}
   % \vspace{-0.1cm}
		\label{fig:instantaneous-violation}
        \vspace{-0.3cm}
\end{figure}

\begin{figure}[t!]
   % \vspace{-0.2cm}
	\centering
\includegraphics[width=\linewidth,height=\textheight,keepaspectratio]{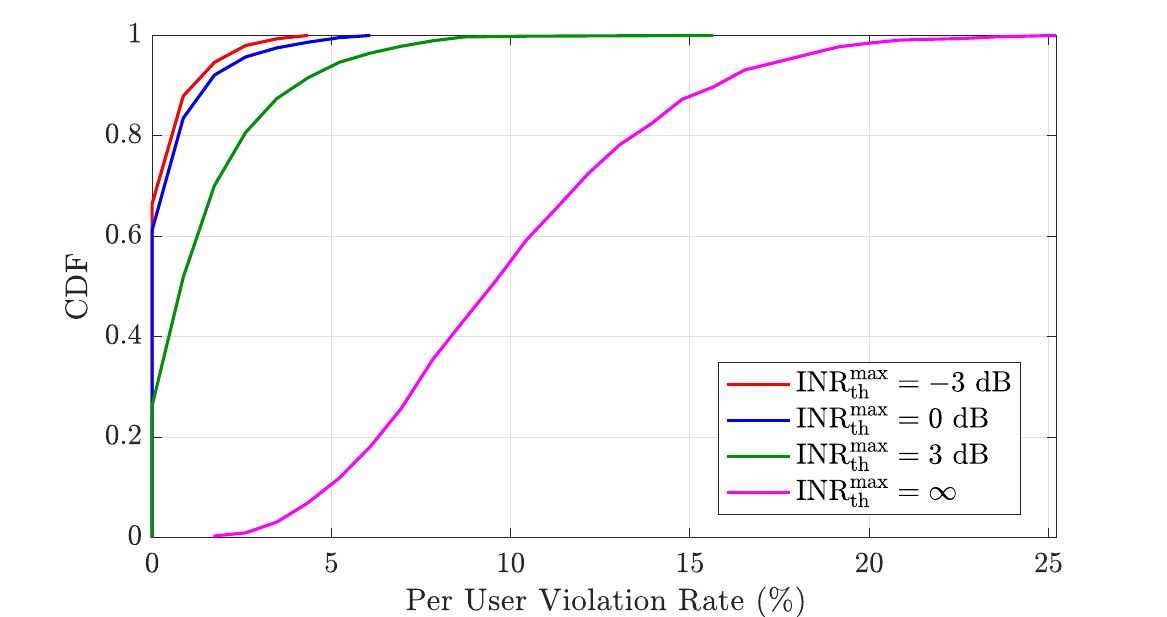}
	\caption{The \gcdf of violation rate per primary user (across time) for various absolute thresholds $\minrthmax$, where $\tminrth = -6$~dB, $\Th = 15$ sec, and $\Tw = 10$ sec.}
		\label{fig:cdf-per-user-violation}
        \vspace{-0.3cm}
\end{figure}

In \figref{fig:instantaneous-violation}, we plot the instantaneous violation rate for various $\minrthmax$ across the same 60-sec period of \figref{fig:inst_inr_max}.
We can see that for $\minrthmax = \infty$, around $10$--$15$\% of users consistently see interference that exceeds $\tminrth = -6$~dB. 
This violation rate exceeds $25$\% at its peak and dips below $5$\% at its lowest.
When setting $\minrthmax = 3$~dB, the violation rate drastically drops to below $5$\%  the vast majority of the time.
This trend continues as $\minrthmax$ is decreased further, intuitively reaching $0$\% when $\minrthmax = \tminrth = -6$~dB.

Further exploring violation, let us examine how frequently each individual primary user experiences violation.
We define the per-user violation rate as the fraction of time in which a primary user's \ginr exceeds $\tminrth$.
\figref{fig:cdf-per-user-violation} shows the \gcdf of the per-user violation rate across all primary users for various absolute thresholds $\minrthmax$, given $\tminrth = -6$~dB. 
With $\minrthmax \leq 3$~dB, we see fairly encouraging results: over 90\% of users experience violations less than 5\% of the time. 
For $\minrthmax = \infty$, all users see violations much more frequently. % s around half of users incur violations at least around 10\% of the time.
Still, it is encouraging to see that more than half of users see violations less than 10\% of the time. 
Moreover, we can see that a worst-case user sees violations at most 25\% of the time.

To shed light on utilization along with violation, \tabref{tab:utilization_violation_table} summarizes both for various time-average thresholds $\minrth$ and absolute thresholds $\minrthmax$. 
Let us first recognize that, when $\tminrth = \minrthmax = -12.2$~dB, which corresponds to a conventional strict protection constraint as described in \eqref{eq:strict_ipr}, there is no violation, but this leads to an average utilization of merely $3.47$\%.
% This further illustrates that 
Similarly, when $\tminrth = \minrthmax = -6$~dB, utilization is only $22.16$\%.
The average violation naturally increases as $\minrthmax$ is relaxed but only modestly so. 
Average utilization, on the other hand, increases substantially to over $80$\% with $\tminrth = -12.2$~dB and to $100$\% for $\tminrth = -6$~dB.
% The ability to trade minor increases in violation rate for major increases in utilization rate is the key advantage of our proposed protection constraint, compared to a conventional strict constraint.

% Achieving a sensible balance between the average \ginr threshold and absolute threshold is needed for effective coexistence 
%\vspace{-0.16cm}
\begin{takeaway}[Trading off minor increases in violation rate for major increases in utilization rate is the key advantage of our proposed protection constraint]
% For extremely low $\tminrth$ values, the secondary system must make significant sacrifices to comply with the constraint, leaving nearly $50 \%$ of its resources underutilized. 
With a conventional strict constraint, the secondary system must make substantial sacrifices in utilization to meet extremely low \ginr thresholds across all users throughout entire handover periods.
With our approach, however, short-lived spikes in \ginr are tolerated in order to enhance secondary system utilization, while still ensuring a certain level of protection is maintained at all users throughout handover periods.
This balance is crucial for effective coexistence. 
 %\vspace{-0.15cm}
\end{takeaway}

% \begin{figure}[t!]
%     \vspace{-0.2cm}
% 	\centering
% \includegraphics[width=\linewidth,height=\textheight,keepaspectratio]{nn_fig/cdf_per_user_violation_r1}
% 	\caption{The \gcdf of violation rate per primary user (across time) for various absolute thresholds $\minrthmax$, where $\tminrth = -6$~dB, $\Th = 15$ sec, and $\Tw = 10$ sec.}
% 		\label{fig:cdf-per-user-violation}
%         \vspace{-0.3cm}
% \end{figure}

\begin{table}
    \vspace{-0.2cm}
	\caption{Utilization and Violation ($\%$)}
	\centering
	\label{tab:}
	\begin{tabular}{|c|c|c||c|c| }
            \hline
            & \multicolumn{2}{|c||}{$\tminrth = -12.2$~dB} & \multicolumn{2}{c|}{$\tminrth = -6$~dB} \\
		\hline
		\minrthmax& Utilization & Violation & Utilization & Violation \\
		\hline
        $-12.2$~dB & 3.47 & 0 & N/A & N/A \\
        \hline
	$-6$~dB & 22.16 & 1.34 & 22.16 & 0 \\
        \hline
        $-3$~dB & 53.44 & 3.70 & 56.45&0.47\\
	\hline 
	$0$~dB & 66.05 & 4.55 & 73.75 & 0.63 \\
	\hline 
	$3$~dB & 84.76 & 10.09 & 93.92 & 1.79 \\
	\hline
	$\infty$& 87.41 & 13.07 &100 & 10.20 \\
	\hline
	\end{tabular}
	\label{tab:utilization_violation_table}
	\vspace{-0.35cm}
\end{table}

% When $\minrthmax = -6$~dB, letting $\tminrth = -12.2$~dB

% \begin{figure}
%     \vspace{-0.25cm}
% 	\centering
% 	\includegraphics[width=\linewidth,height=0.25\textheight,keepaspectratio]{nn_fig/ho_freq_histogram1}
% 	\caption{Histogram of actual handover periods for various INR thresholds $\tminrth = \tminrth^\mathrm{max}$ with secondary satellite selection updated at every slot. 
% Over $80\%$ of associations last for only one slot duration $0.1$ second. }
% 		\label{fig:histogram}
%         \vspace{-0.35cm}
% \end{figure}

\comment{
Thus far, we have assumed an ideal scenario where handovers may occur at every slot, $0.1$ seconds, which is not practical in real-world settings. Before examining the impact of the handover period on the secondary satellite selection, we first analyze how long a satellite-to-cluster association lasts when the secondary system is allowed to select satellites at every time slot. 
Surprisingly, satellite-to-cluster associations tend to be short-lived; more than $80 \%$ of associations persist for only one slot ($0.1$ second), as shown in \figref{fig:histogram}.
The optimal satellite for one cluster at one instance may become optimal for another at the next instance and then revert to optimal for the original cluster in subsequent instances, which highlights the dynamic nature of \gsinr environments in dense \leo satellite systems, driven by the rapid motion of satellites orbiting the globe.
These frequent handover statistics underscore the need to regulate the handover period, which can be achieved by introducing a regularization component in the objective of the problem \eqref{eq:objective} or systematically defining handover criteria as an operation parameter. In the following sections, we closely examine the impact of the handover period and the \ginr averaging window size on system performance.
}

\begin{figure*}
    \vspace{-0.2cm}
    \centering
    \subfloat[Average violation rate.]{\includegraphics[width=0.5\linewidth,height=\textheight,keepaspectratio]{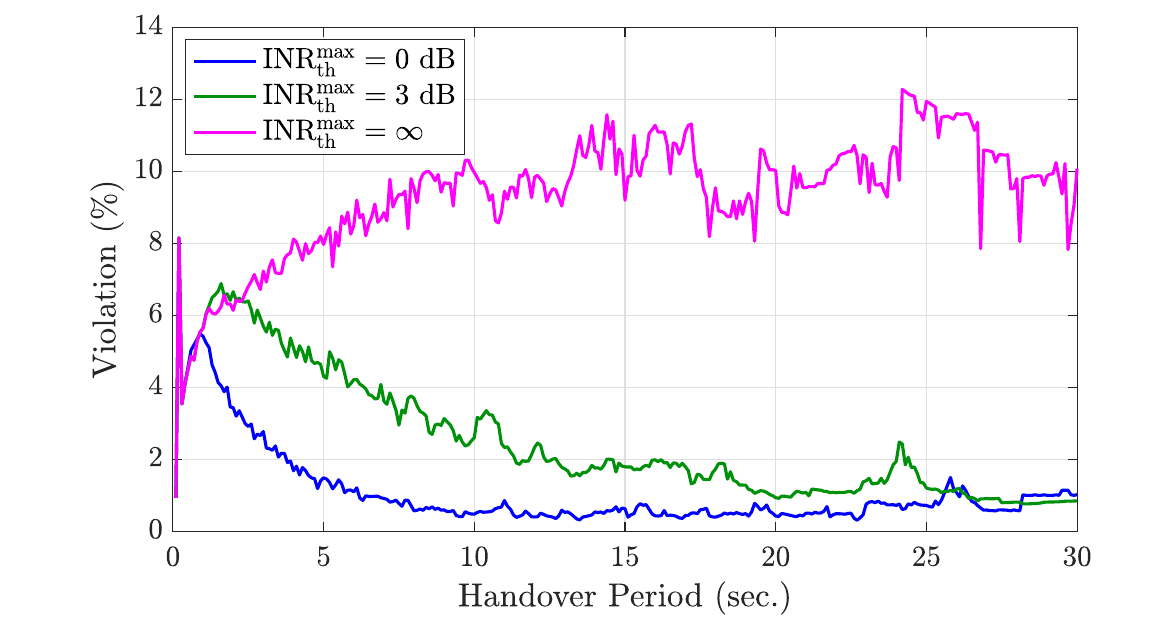}\label{fig:handover_violation}}
    %\quad
    \subfloat[Average secondary system utilization.]{\includegraphics[width=0.5\linewidth,height=\textheight,keepaspectratio]{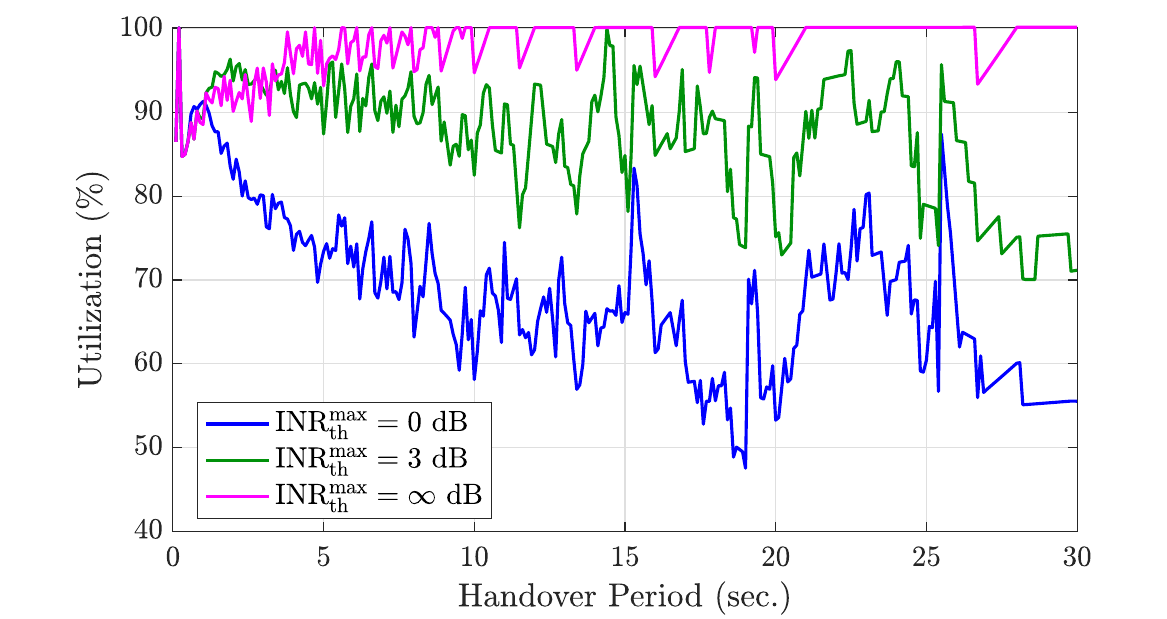}\label{fig:handover_utilization}}
    \caption{Average violation and utilization rates as a function of handover period $\Th$ for various $\minrthmax$, when $\tminrth=6$~dB and $\Tw = 10$ sec. When $\minrthmax$ is high, increasing $\Th$ tends to increase utilization, since it provides longer durations to average out spikes in \ginr, but this leads to a higher violation rate. When $\minrthmax$ is low, increasing $\Th$ tends to decrease utilization since the secondary system cannot meet the protection without putting some clusters into outage; this naturally decreases the violation rate.}
    % Allowing higher \ginr the secondary system to inflict the primary users temporarily while maintaining the reference threshold $\tminrth=-6$ dB can increase utilization of the secondary system and maintain lower violation.
    \label{fig:handover_period}
    \vspace{-0.4cm}
\end{figure*}

Let us now examine \figref{fig:handover_violation}, which depicts average violation rate as a function of the handover period $\Th$, for $\tminrth=-6$~dB and various $\minrthmax$.
We can see that the violation rate is generally lower for a stricter absolute protection constraint $\minrthmax$, but we can observe two noteworthy trends beyond this.
% We can also observe two trends.
First, when $\minrthmax = \infty$, the violation rate tends to increase with handover period $\Th$.
This can be explained by the fact that a longer handover period allows the system to tolerate (average out) high spikes in \ginr while still meeting the time-average protection constraint. 
Even when $\minrthmax = \infty$, the violation rate caps at about $12\%$.
The second trend can be stated as, when $\minrthmax$ is lower, violation rate tends to decrease as the handover period is increased.
This is because, for strict absolute thresholds $\minrthmax$, it is more difficult for the secondary satellite system to meet the constraint; this exacerbates as the handover period is increased, since the same association is used throughout the entire handover period.
Consequently, the secondary system is more often incapable of serving some clusters when $\minrthmax$ is low, which naturally reduces the violation rate. 

This can be confirmed from \figref{fig:handover_utilization}, which shows the corresponding secondary system utilization as a function of handover period.
% From \figref{fig:handover_utilization}, 
We can see that the secondary system utilization is heavily impacted by the handover period for strict $\minrthmax$, as utilization can drop to around $50\%$ for a handover period of $\Th \approx 18$ sec.
Again, this is because the secondary system is incapable of meeting such a strict absolute constraint across all users throughout long handover periods and is forced to leave certain clusters in outage.
When $\minrthmax$ is increased, utilization increases thanks to the relaxed protection constraint.
However, even in the most relaxed case when $\minrthmax = \infty$, there are select cases where the time-average constraint cannot be met.
Furthermore, one can notice that when the handover period $\Th$ is made small (which relaxes the absolute constraint), the utilization still falls short of 100\% and is in fact not maximized.
This can be best explained by the fact that a small $\Th$ results in a more restrictive time-average constraint, as a shorter handover period offers less time to average out spikes in interference. 
When $\Th$ is increased, more frequent and/or higher spikes in interference can be tolerated, as there is more time to average them out and meet the time-average constraint $\tminrth$.
Thus, to maximize secondary system utilization under a given $\tminrth$, $\minrthmax$ and $\Th$ should both be high.

% \vspace{-0.17cm}
\begin{takeaway}[Optimizing coexistence entails careful selection of a number of parameters]
Given the complexity of this interference scenario, it is difficult to state a one-size-fits-all solution.
With our proposed framework, however, multiple parameters can be tuned to reliably arrive at a satellite selection mechanism which meets an acceptable violation rate while maximizing utilization.
Regulators will play a pivotal role in defining violation rates and worst-case \ginr that are acceptable in this context.
Beyond this, the rate at which the secondary system can perform handovers will also have decisive impact on its ability to coexist with the primary system.
%\vspace{-0.12cm}
\end{takeaway}

% \textbf{Heatmaps:}
% Thus far, the time window has been set to $\Tw = 10$ seconds throughout our results.
% We now extend this evaluation of utilization and violation rate for many handover periods $\Th$ and time windows $\Tw$ in \figref{fig:heatmap1}, fixing $\tminrth = -6$~dB and $\minrthmax = 3$~dB.
% \ipr{Circle back to this.}

% ---

% Let us begin by examining \figref{fig:handover_utilization}, which depicts the secondary system utilization for a various $\minrthmax$ as a function of the handover period $\Th$.
% First consider a somewhat strict $\minrthmax = 0$~dB, which shows 

% Coexistence performance is assessed using two metrics: (i) the violation rate, defined as the fraction of primary users whose instantaneous \ginr exceeding the average \ginr threshold \tminrth due to the secondary system, indicating compliance with the interference constraints; and (ii) secondary system utilization, measured as the fraction of ground users actively served by secondary satellites. 

\comment{
\begin{figure*}
	\centering
	\subfloat[Secondary system utilization]{\includegraphics[width=\linewidth,height=0.25\textheight,keepaspectratio]
    {nn_fig/utilization_w_vs_h_r1}%{nn_fig/utilization_w_vs_h(0.5)_offset_r1}
		\label{fig:heatmap_utilization}}
	\qquad\qquad
	%	\pause 
	\subfloat[Violation of $-12.2$ dB.  	]{\includegraphics[width=\linewidth,height=0.25\textheight,keepaspectratio]
    {nn_fig/violation_w_vs_h_r1}%{nn_fig/violation_w_vs_h(0.5)_offset_r1}
		\label{fig:heatmap_violation}} \\
 %    \subfloat[Maximum value of Primary INR	]{\includegraphics[width=\linewidth,height=0.25\textheight,keepaspectratio]{nn_fig/max_inr_over_H_per_inr_max(w=1)}}
	\caption{%Heatmaps of utilization and violation of $\tminrth=-6$ dB as a function of handover period and averaging window size with $\minrth^{\mathrm{max}}$ to be infinite. The higher utilization of the secondary system tends to lead higher violation of the reference threshold, but for some combinations of $\Th$ and $\Tw$ can lead higher utilization with lower violation.
    }
	\label{fig:heatmap1}
\end{figure*}

\comment{
\begin{figure*}
	\centering
	\subfloat[95 percentile INR]{\includegraphics[width=\linewidth,height=0.25\textheight,keepaspectratio]{nn_fig/percentile(95)_w_vs_h(0.5)_offset_r1}
		\label{fig:heatmap_95}}
	\qquad\qquad
	\subfloat[Maximum INR]{\includegraphics[width=\linewidth,height=0.25\textheight,keepaspectratio]{nn_fig/max_inr_w_vs_h(0.5)_offset_r1}
		\label{fig:heatmap_max}} \\ 
 %    \subfloat[Maximum value of Primary INR	]{\includegraphics[width=\linewidth,height=0.25\textheight,keepaspectratio]{nn_fig/max_inr_over_H_per_inr_max(w=1)}}
	\caption{Heatmaps of $95$-percentile and realized maximum \ginr of primary users as a function of handover period and averaging window size when $\tminrth=-12.2$ dB and $\minrth^\mathrm{max}= \inf$.   
    As $\Th$ and $\Tw$ increase, the maximum \ginr and $95$-percentile \ginr of the primary \ginr also increase but saturated.}
	\label{fig:heatmap2}
\end{figure*}
}
In \figref{fig:heatmap1}, we present the heatmap of utilization and violation of $\tminrth=-12.2$~dB as a function of $\Th$ and $\Tw$, with $\minrth^{\mathrm{max}}$ set to infinity. 
% \edit{The frame time of the secondary system is offset by $n_0=2$ seconds from that of the primary system. }
 \figref{fig:heatmap_utilization} shows that increasing $\Th$ and $\Tw$, which results in relaxation of the \ginr protection constraint, leads to higher utilization of the secondary system.  
 While higher utilization generally corresponds to a higher violation, we are particularly interested in whether the satellite selection algorithm can yield higher utilization with reasonably acceptable low violation. 
 It is encouraging to observe lower violation rates of $-12.2$~dB in \figref{fig:heatmap_violation} while achieving higher secondary system utilization for some combinations of $\Th$ and $\Tw$ values shown in \figref{fig:heatmap_utilization}.
 
 Examining the $x$-axis of \figref{fig:heatmap_violation}, which represents the handover period $T_\mathrm{h}$, the secondary system can maintain relatively low violation rates up to specific values of $T_\mathrm{h}$, approximately between $18$ to $22$ seconds. 
This can be because the secondary system selects serving satellites considering the inflicted \ginr onto primary users until the next handover instance. 
However, beyond these values of $T_\mathrm{h}$,
the violation rate increases due to changes in the primary system's serving satellites, which occur every $15$ seconds in this study. 
As $T_\mathrm{h}$ increases further, minimizing interference on primary users becomes increasingly difficult, as the interference environment changes significantly between two completely different sets of primary serving satellites, leading to higher violation rates.
}

\comment{
\figref{fig:heatmap2} illustrates the heatmap of $95$-percentile and maximum primary user \ginr as a function of $T_\mathrm{w}$ and $T_\mathrm{h}$. 
Increasing $T_\mathrm{w}$, and $T_\mathrm{h}$ leads to relaxation of the interference protection constraint for the secondary system: as $T_\mathrm{w}$, and $T_\mathrm{h}$ increase, the maximum realized \ginr values of the primary \ginr also increase. 
In all cases, the maximum values of \ginr range from $-10$ to $12$~dB, which is notable given that $\tminrth^{\mathrm{max}}$ is set to infinity while $\tminrth=-12.2$~dB. 
While the maximum values of \ginr may appear high, the $95$-percentile values of the primary \ginr are significantly more restrictive, ranging from $-14$ to $-5$~dB. This shows that the secondary satellite selection algorithm effectively controls the \ginr inflicted onto primary users depending on the $\timew$, $\timeho$, \tminrth, and $\tminrth^{\mathrm{max}}$. 
These parameters, which can be system-specific or regulatory, are adjustable within the selection algorithm to ensure compliance and optimal performance.
}

% \vspace{-0.25cm}
%\section{Learning Primary Satellite-to-Cluster Associations}
\section{Coexistence Under Uncertainty\\about the Primary System}
\label{sec:nn}

Until now, we have implicitly assumed that the secondary system has perfect knowledge of the primary system association at any given time. 
This assumption may be sound in cases where the two systems communicate with one another (or through a third party, such as a regulator), but this is not expected nor mandated in today's \leo landscape.
Motivated by this, this section presents and assesses schemes which use DL to learn the primary system's association policy. 
More specifically, we propose using secondary ground users to measure the received power from primary satellites, which can then be used to infer the primary serving satellites over time in a given area.
These inferred primary serving satellites are then used to train a centralized DL model, which can then be used to forecast the primary satellite which serves a given location at any point in the future, based solely on the primary satellite locations---which is publicly available.

\comment{
Although satellites in an orbital plane follow periodic trajectories around the globe based on their orbital parameters, regularly passing over specific ground locations, accurately identifying the exact pattern of overhead satellite locations in dense \leo constellations remains challenging. The high satellite density and dynamic orbital configurations result in complex satellite visibility patterns, making it difficult to predict associations using simple deterministic models. This further justifies the need for a learning-based approach to infer satellite-to-cluster associations effectively.
}

%\subsection{Learning Algorithms}

\comment{
\begin{figure}
    \vspace{-0.25cm}
	\centering
	\includegraphics[width=\linewidth,height=0.24\textheight,keepaspectratio]{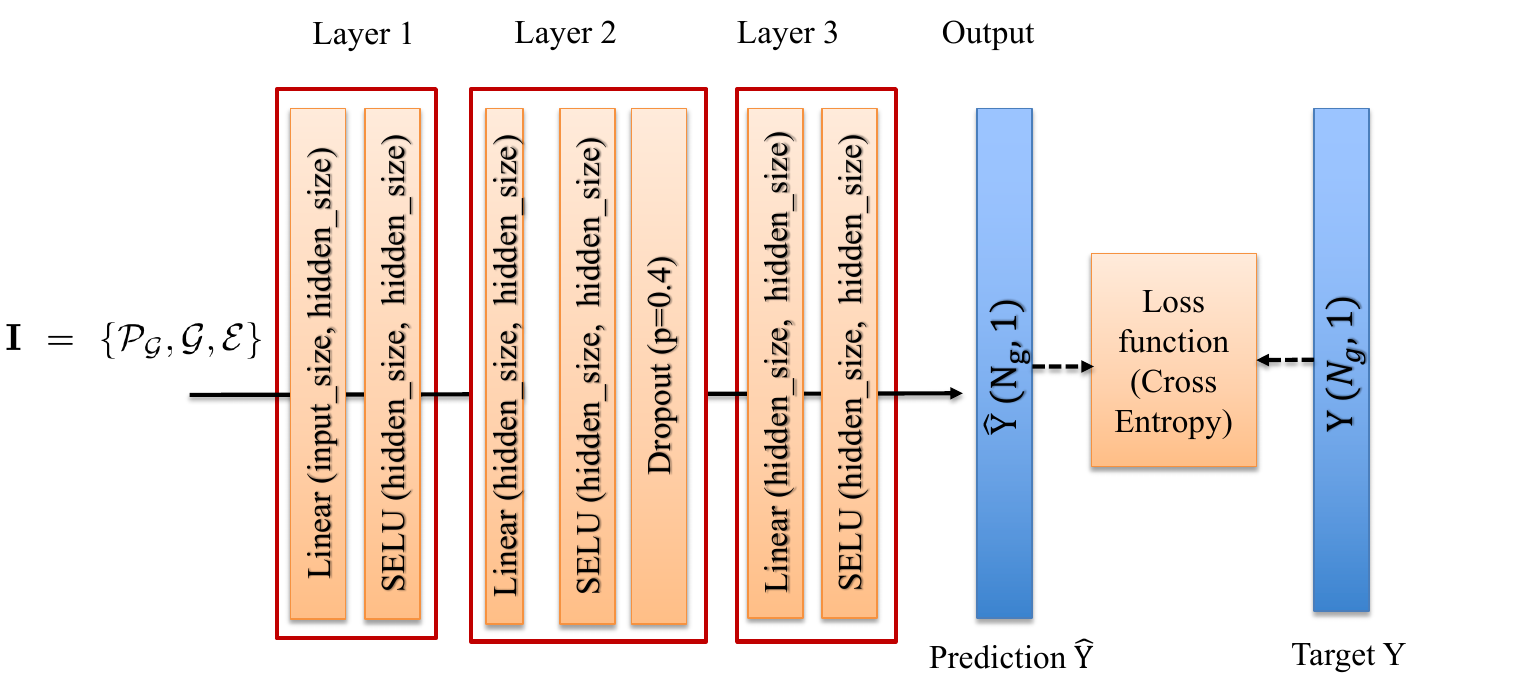}
	\caption{Inputs and targets of our three layer MLP model.}
    	\label{fig:mlp}
        \vspace{-0.15cm}
\end{figure}
}

% We expect that \gnn can learn the primary system’s satellite-to-ground cluster association policies by capturing spatial relationships and temporal patterns within the data. Once trained, the \nn can predict---hopefully accurately---which satellite will serve a specific ground location at any given time using publicly available primary satellite location information. %, assuming that the primary system maintains consistent association policies within the considered area.

\setcounter{equation}{22}
Let the input of our DL model be $\bI = \braces{\mathcal{P}^t_\mathcal{G}, \mathcal{G}, \mathcal{E}^t}$, comprised of the primary available satellite locations at time $t$, the locations of all cluster centers, and the elevation angles of all available satellites with respect to each cluster at time $t$. 
% with the input size $(3\numa +3\numg+\numg\numa)$. 
The output of the DL model $\hat{\bY} = f(\bI;\boldsymbol{\theta})\in \mathbb{R}^{\numg}$ represents the index of the estimated primary serving satellite for each cluster, where $f(\cdot;\boldsymbol{\theta})$ is a learnable function defined by the DL model's parameters $\boldsymbol{\theta}$. 
% , while $\bY$ denotes the true primary serving satellite indices. 
To compare the effectiveness of different DL models, we evaluated a single-layer (1L) perceptron, a three-layer (3L) multilayer perceptron (MLP), and an attention-based network. 
For the perceptron models, we use the scaled exponential linear unit (SELU) \cite{selu} as the activation function to enable self-normalization with a hidden size of 2048. 
Training data was generated by simulating the primary system’s satellite-to-cluster associations using Starlink’s constellation over the considered geographical area in \figref{fig:geo_map}, using all of the same simulation parameters described in the previous section. % To enrich the dataset, we repeat this proc

\tabref{tab:mlp} presents the test accuracy of the considered DL models for each handover policy. % : HE and MCT. 
Note that the model is not provided any explicit information about the primary system's handover policy, as this is what it must learn. 
Accuracy is computed as the percentage of clusters (averaged over time) for which the model correctly predicts their serving primary satellites. %  across all considered clusters. 
% showing slightly better in learning MCT policy. 
The results indicate that the three DL models have difficulty in precisely identifying the primary serving satellites, achieving accuracies that range from about $51$--$67\%$.
This suggests that learning even fairly simple association policies like HE and MCT is a more complex task than it may appear. %  than initially anticipated. 
Tackling this problem more extensively in dedicated work would thus make a valuable contribution.
\comment{ \red{Delete?} One contributing factor is that a satellite with the highest elevation (or longest visible time) for one cluster may also have the highest elevation (or longest visible time) for other clusters. Additionally, the set of overhead satellites varies across clusters, further complicating the learning process and reducing prediction accuracy.
 }

On the other hand, given that the number of available satellites ranges from approximately $20$ to $25$ at each time instance, our learning algorithm shows promising results by correctly identifying the serving satellite for at least five out of the ten clusters. 
Although ``$5$ out of $10$ correct'' may seem like modest accuracy, the ample combinatorial possibilities in assigning a unique satellite to each of the 10 clusters indicate that this accuracy is still noteworthy. 
Additionally, the table provides accuracy metrics when predicting the top three or top five potential serving satellites, with performance for top-5 easily exceeding $90\%$. 
The implications of the DL models' accuracy will be further evaluated in the context of primary and secondary satellite system coexistence in the following.

\begin{table}
    %\vspace{-0.12cm}
	\caption{Test Data Accuracy of Various DL Models ($\%$)}
     \vspace{-0.12cm}
	\centering
	\label{tab:mlp}
	\begin{tabular}{|c|c|c|c|c|}
		\hline
		 Policy&DL Model & Top-$1$& Top-$3$ & Top-$5$\\
        % \hline
        % \multicolumn{4}{|c|}{HE handover policy applied}\\
		\hline
		\multirow{3}{*}{HE}& 1L Perceptron & $57.66$ &$88.21$ & $95.26$\\
        %62.58 &90.31  & 96.10 \\
        \cline{2-5}
		&3L MLP& $55.75 $& $86.95$ & $95.19$ \\ 
        %62.11  & 90.11 & 94.16\\
		\cline{2-5}
		&Attention &$51.21$ & $83.52$ & $92.27$\\
        %&62.27 & 90.24 & 95.77\\
		\hline
        \multirow{3}{*}{MCT}& 1L Perceptron & $60.86 $& $88.83$  & $96.07$ \\
        \cline{2-5}
		&3L MLP& $67.12$ & $91.34$ & $97.01$\\
        % 47.35, 83.82, 93.96
		\cline{2-5}
		&Attention &$57.94$ & $86.72$ & $94.78$\\
        % 41.81, 84.35, 93.93
		\hline
	\end{tabular}
    \vspace{-0.35cm}
\end{table} 

\begin{table}[t]
    %\vspace{-0.12cm}
	\caption{Violation/Utilization Rate (each in $\%$) of $\tminrth$\\($\Tw=10$ sec, $\Th = 15$ sec, $\minrmax = \infty$) }
    \vspace{-0.12cm}
	\centering
	\label{tab:}
	\begin{tabular}{|c|c|c|c|c| }
    % ans =

		\hline
		\tminrth & True Sat.&  Top-$1$ &Top-$3$ & Top-$5$ \\
		\hline
		$-12.2$~dB &$13.07$/$87.41$&  $15.74$/$42.53$ &    $2.13$/$26.04$  & $2.24$/$14.03$\\
        \hline
         $-10$~dB &$12.91$/$92.04$ & $13.63$/$87.19$  &  $2.63$/$32.19$&   $1.84$/$20.03$\\
        \hline
        $ -6$~dB & $10.22$/$100$ & $ 10.71$/$98.53$  &  $3.81$/$46.02$ &  $0.71$/$24.04$\\
      \hline
        $ 0$~dB & $1.71$/$100$ & $1.83$/$100$ &  $ 1.92$/$98.01$  &  $1.34$/$80.04$\\
		\hline
	\end{tabular}
	\label{tab:violation_utilization_nn}
	\vspace{-0.3cm}
\end{table}

The coexistence performance of the secondary system, based on the inferred primary system's satellite-to-cluster association using a single-layer (1L) perceptron, is presented in Table~\ref{tab:violation_utilization_nn}, which reports utilization and violation metrics for various $\tminrth$ values.
Despite the modest prediction accuracy being below $60\%$ (Table~\ref{tab:mlp}), both utilization and violation metrics remain close to those obtained using true primary association data, except in the case of $\tminrth=-12.2$~dB. 
Specifically, when $\tminrth=-10$~dB, the utilization of the secondary system under the top-1 estimate is approximately $5\%$ lower than that obtained with true data, while the violation increases by less than $2\%$.
These results indicate that, under a strict average threshold of $-12.2$~dB, a single violation due to an incorrect primary satellite estimate can disproportionately impact the association, highlighting an opportunity for further improvements of primary satellite prediction. 
However, with a more relaxed constraint, the inferred primary association data using the simple 1L perceptron enables performance comparable to that achieved with true data. 

On the other hand, despite top-3 and top-5 estimates exhibiting a high prediction accuracy, notable reduction in utilization and violation is observed when the secondary system relies on the top-3 or top-5 estimates. 
This indicates that the secondary system overly protects the primary system and excessively refrain from assigning satellites to its clusters. 

% However, the violation of secondary satellite selection decreases a lot when using the estimated primary associations of top three or five. % especially for very low $\tminrth$ values. 
% Similarly, a notable reduction in utilization is observed when the secondary system relies on these estimates. This reduction in utilization explains why the violation rate for the Top-$K$ ($K>1$) estimated serving satellites is lower than that for the true data; the secondary system overly protect the primary system and excessively refrain from assigning  serving satellites to its clusters. 

% maybe we don't need to have top-K  results
\comment{
In particular, when using the Top-$K$ ($K>1$) estimated primary satellites as potential serving satellites for each cluster, we slightly modify our secondary satellite selection algorithm to protect all estimated  satellites, for example, the potential inflicted \ginr onto primary users should not exceed the \ginr threshold 
\begin{align}
&\max_{\sfu\in\mathcal{U}} \, \sum_{m=1}^{\nums} \sum_{n=1}^{\numg}  x_{m,n}^t  	\tminrTht(\sfu, \vp^{(k)}_\sfu; \vs_m, \vg_n) \leq \tminrtheff^{t}
\end{align}
for all $k\in\braces{1,\dots, K}$ where $\vp^k_\sfu$ represent the $k$-th estimated serving satellite of primary user $\sfu$. 
}

\comment{
It should also be noted that the secondary system only utilizes $88\%$ when $\tminrth=-12.2$~dB, even with complete knowledge of the primary serving satellites. Combined with the results from the Top-$K$ estimates, this suggests that as the number of protected primary satellites increases, the secondary system adopts a more conservative approach, often refraining from serving ground clusters. While this strategy effectively reduces inflicted \ginr, it comes at the cost of significantly lower utilization. This effect becomes even more pronounced under stricter \ginr thresholds, highlighting the inherent trade-off between interference control and system utilization.
}

\begin{takeaway}[Leveraging DL models can facilitate the coexistence of dense \leo satellite systems] 
The primary system's satellite-to-cluster association can be estimated using a DL model by capturing spatial relationships and temporal patterns within the data. 
These patterns are influenced by satellite locations, handover policies, and cluster-specific priorities.
A simple single-layer (1L) perceptron can achieve a modest accuracy in estimating this association. While high learning accuracy is not strictly necessary, even a moderate level of accuracy can effectively support the secondary system's operation under a relaxed average \ginr constraint. However, under a strict constraint, this modest accuracy can significantly impact performance, emphasizing the need for more sophisticated learning models. %  or improved inference strategies.
\end{takeaway}

\vspace{-0.2cm}
\section{Conclusion and Future Directions} \label{sec:conclusion}

This work investigates the in-band coexistence of two dense LEO satellite communication systems through the lens of two leading commercial entities: SpaceX's Starlink as the primary system and Amazon's Project Kuiper as the secondary system. 
To enable coexistence, we propose a practical mechanism that strategically selects secondary serving satellites to maximize the secondary system capacity while guaranteeing protection to primary ground users.
% meeting the interference protection constraint. 
In doing so, we formulate a novel interference protection constraint that combines an absolute interference threshold and a time-averaged threshold, which provides flexibility in trading off interference violation rate for system utilization.
We use Lagrangian relaxation and subgradient methods to solve our satellite selection problem and thoroughly evaluate performance through high-fidelity simulation based on public FCC filings and technical specifications.  
% The proposed algorithm is validated through high-fidelity simulation that accounts for multiple orbiting satellites from both systems over a broad ground area.

Results show that independent operation of Starlink and Kuiper leads to prohibitively high interference, motivating the need for mechanisms such as ours that explicitly ensure protection.
However, we observe that a conventional absolute protection constraint leads to either under-utilization of the secondary system or high interference at the primary users. 
With our proposed scheme and novel protection constraints, however, utilization and protection can be better balanced by tolerating infrequent, short-lived spikes in interference. 
This allows the primary and secondary systems to both enjoy high capacity while coexisting alongside one another.
Finally, we demonstrated the potential use of DL techniques to infer the primary satellite associations at any given time based purely on the knowledge of overhead satellites, which can then be used by our proposed scheme to find near-optimal secondary satellite associations to enable coexistence.

A number of valuable directions are motivated by this work, including more extensively exploring DL techniques to improve the estimation of the primary system's satellite-to-cluster association. 
Additionally, it would be valuable to consider cases where there is limited communication/coordination between the primary and secondary systems. 
Another promising direction is to investigate the coexistence of three or more satellite constellations, extending our analysis to more complex multi-system scenarios.

%\input{sec-appendix.tex}

%\vspace{-0.2cm}

\section*{Acknowledgments}

We thank Arun Ghosh, Michael Hicks, Anil Rao, and Vikram Chandrasekhar from Amazon's Project Kuiper for their %valuable 
discussions and feedback on our methodology and results.
%\vspace{-0.22cm}

\bibliographystyle{bibtex/IEEEtran}
\bibliography{bibtex/IEEEabrv, refs}
%\bibliography{refs}

\end{document}